\documentclass[draftcls,onecolumn,12pt]{IEEEtran}
\usepackage[utf8]{inputenc}
\usepackage{amsfonts,amsmath,mathrsfs,bm}
\usepackage{algorithm}
\usepackage{algpseudocode}
\usepackage{amssymb}
\usepackage{balance}
\usepackage{cite}
\usepackage[font=footnotesize]{caption}
\usepackage[font=footnotesize]{subcaption}
\usepackage{graphicx}
\usepackage{standalone}
\usepackage{tikz}
\usepackage{tikz-3dplot}
\usepackage[nolist]{acronym}

\begin{acronym}[ACRONYM]
\acro{AI}{artificial intelligence}
\acro{AoA}{angle-of-arrival}
\acro{AoD}{angle-of-departure}
\acro{BS}{base station}
\acro{BP}{belief propagation}
\acro{CDF}{cumulative density function}
\acro{CFO}{carrier frequency offset}
\acro{CRB}{Cram\'er-Rao bound}
\acro{DA}{data association}
\acro{D-MIMO}{distributed multiple-input multiple-output}
\acro{DL}{downlink}
\acro{EM}{electromagnetic}
\acro{FIM}{Fisher information matrix}
\acro{GDOP}{geometric dilution of precision}
\acro{GNSS}{global navigation satellite system}
\acro{GPS}{global positioning system}
\acro{IP}{incidence point}
\acro{IQ}{in-phase and quadrature}
\acro{ISAC}{integrated sensing and communication}
\acro{ICI}{inter-carrier interference}
\acro{JCS}{Joint Communication and Sensing}
\acro{JRC}{joint radar and communication}
\acro{JRC2LS}{joint radar communication, computation, localization, and sensing}
\acro{IMU}{inertial measurement unit}
\acro{IOO}{indoor open office}
\acro{IoT}{Internet of Things}
\acro{IRN}{infrastructure reference node}
\acro{KPI}{key performance indicator}
\acro{LoS}{line-of-sight}
\acro{LS}{least-squares}
\acro{MCRB}{misspecified Cram\'er-Rao bound}
\acro{MIMO}{multiple-input multiple-output}
\acro{ML}{maximum likelihood}
\acro{mmWave}{millimeter-wave}
\acro{NLoS}{non-line-of-sight}
\acro{NR}{new radio}
\acro{OFDM}{orthogonal frequency-division multiplexing}
\acro{OTFS}{orthogonal time-frequency-space}
\acro{OEB}{orientation error bound}
\acro{PEB}{position error bound}
\acro{VEB}{velocity error bound}
\acro{PRS}{positioning reference signal}
\acro{QoS}{Quality of Service}
\acro{RAN}{radio access network}
\acro{RAT}{radio access technology}
\acro{RCS}{radar cross section}
\acro{RedCap}{reduced capacity}
\acro{RF}{radio frequency}
\acro{RIS}{reconfigurable intelligent surface}
\acro{RFS}{random finite set}
\acro{RMSE}{root mean squared error}
\acro{RTK}{real-time kinematic}
\acro{RTT}{round-trip-time}
\acro{SLAM}{simultaneous localization and mapping}
\acro{SLAT}{simultaneous localization and tracking}
\acro{SNR}{signal-to-noise ratio}
\acro{ToA}{time-of-arrival}
\acro{TDoA}{time-difference-of-arrival}
\acro{TR}{time-reversal}
\acro{TXRX}[TX/RX]{transmitter/receiver}
\acro{Tx}{transmitter}
\acro{Rx}{receiver}
\acro{UE}{user equipment}
\acro{UL}{uplink}
\acro{UWB}{ultra wideband}
\acro{XL-MIMO}{extra-large MIMO}
\end{acronym}

\title{A Projective Geometric View for 6D Pose Estimation in mmWave MIMO Systems}
\author{Shengqiang Shen~\IEEEmembership{Member,~IEEE}, Henk Wymeersch~\IEEEmembership{Senior Member,~IEEE}
\thanks{Shengqiang Shen is with the School of Information and Control Engineering, China University of Mining and Technology, 221000 Xuzhou, China (e-mail: sshen@cumt.edu.cn). Henk Wymeersch is with the Department
of Electrical Engineering, Chalmers University of Technology, 41258 Gothenburg, Sweden (e-mail: henkw@chalmers.se).}
\thanks{This work was supported, in part, by the European Commission through the H2020 project Hexa-X (Grant Agreement no. 101015956) and by the Swedish Research Council under grant No. 2018-03701.}}

\usepackage[nameinlink,capitalise]{cleveref}

\crefname{equation}{}{}
\crefrangemultiformat{equation}{(#3#1#4--#5#2#6)}{ and~(#3#1#4--#5#2#6)}{, (#3#1#4--#5#2#6)}{ and~(#3#1#4--#5#2#6)}

\begin{document}

\newcommand{\bmsf}[1]{\bm{\mathsf{#1}}}
\newcommand{\vect}[1]{\mathrm{vec}\left({#1}\right)}
\maketitle

\begin{abstract}
Millimeter-wave (mmWave) systems in the 30--300 GHz bands are among the fundamental enabling technologies of 5G and beyond 5G,  providing large bandwidths, not only for high data rate communication, but also for precise positioning services, in support of high accuracy demanding applications such as vehicle positioning.
With the possibility to introduce relatively large arrays on user devices with a small footprint, the ability to determine the user orientation becomes unlocked. The estimation of the full user pose (joint 3D position and 3D orientation) is referred to as 6D localization.  %
Conventionally, the problem of 6D localization using antenna arrays has been considered difficult and was solved through a combination of heuristics and optimization. 
In this paper, we reveal a close connection between the \acp{AoA} and \acp{AoD} and 
the well-studied perspective projection model from computer vision. This connection allows us to solve the  6D localization problem, by adapting state-of-the-art methods from computer vision. More specifically, two problems, namely 6D pose estimation from \acp{AoA} from  multiple single-antenna base stations and 6D SLAM based on single-BS mmWave communication, are first modeled with the perspective projection model, and then solved. 
Numerical simulations show that the proposed estimators operate close to the  theoretical performance bounds. Moreover, the proposed SLAM method is effective even in the absence of the \ac{LoS} path, or knowledge of the \ac{LoS}/\ac{NLoS} condition.
\end{abstract}

\acresetall %
\section{Introduction}
Continuous development of the fifth-generation (5G) network intends to broaden its uses beyond traditional mobile broadband and to enable a key capability of precise positioning, which is expected to be required in a variety of new applications \cite{saily2021positioning}, in particular, for vehicles \cite{5gppp5g}. Among the developments towards 5G Advances are the support of sidelinks, improved integrity, carrier phase positioning, and the support of reduced capacity (RedCap) devices \cite{nikonowicz2022indoor,chen2022standardization}.  
As one of the key enabling components of 5G and beyond 5G, millimeter-wave (mmWave) provides massive bandwidths for high data rates and empowers precise positioning services using cellular technology rather than a separate infrastructure.
The possibility to introduce a relatively large array on user devices in mmWave MIMO systems brings the ability to estimate user orientation, in addition to the user position \cite{chen2022tutorial}. Information of the full user pose  %
not only benefits the performance of communication systems \cite{zhao2018beam}, but also can be used for higher level applications \cite{struye2021millimeter}, particularly driving assistance applications and platooning in intelligent transport systems \cite{Bartoletti21Positioning}. As the external device that is able to provide the pose information is not always available, due to cost and/or size limits. As a result, there has been an increasing interest in the use of antenna arrays for joint position and orientation estimation, i.e., pose estimation, referred to here as 6D localization \cite{shastri2022review}. 

\begin{figure}
	\centering
		\includegraphics[width=0.9\columnwidth]{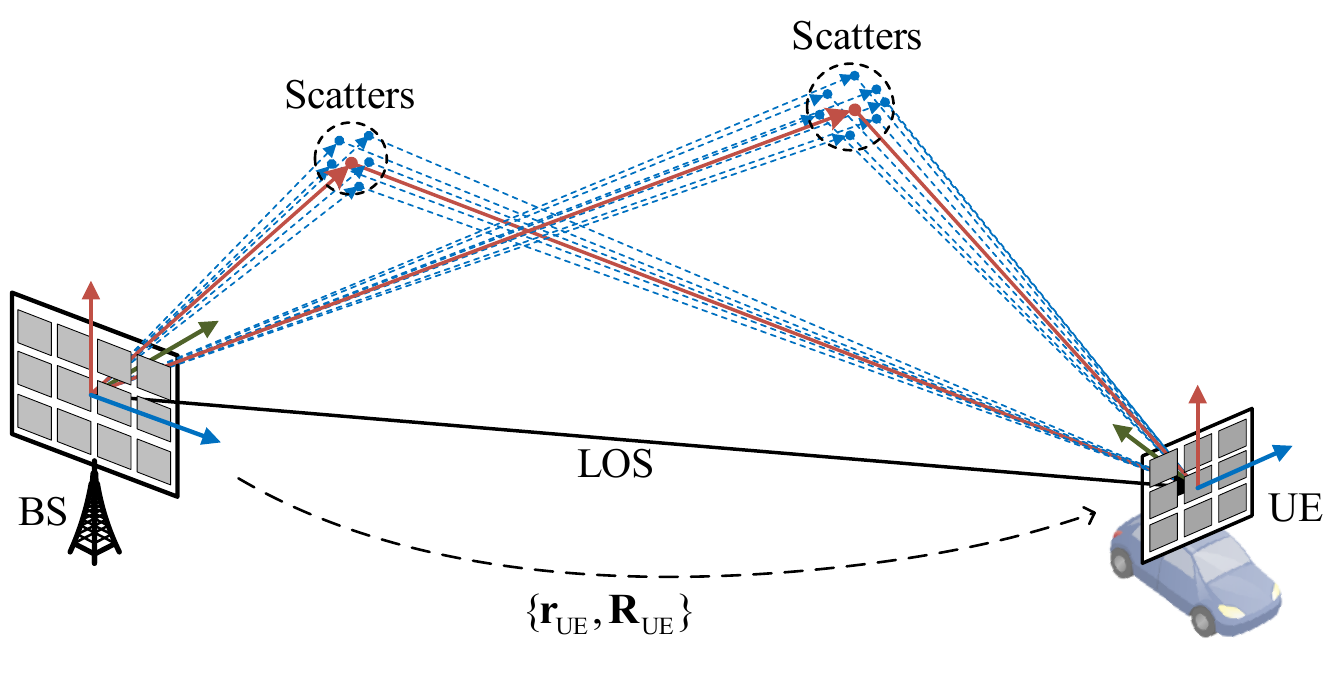}
	\caption{Illustration of 3D MIMO system, comprising a BS with known pose, a UE with unknown pose (location $\mathbf{r}_{\mathrm{UE}}$ and orientation $\mathbf{R}_{\mathrm{UE}}$) in an environment with several scatterers with unknown locations. }
	\label{fig:channelModel} \vspace{-5mm}
\end{figure}

Two system configurations are typically considered for 6D localization, i.e, the multi-BS and the single-BS cases. In the multi-BS case, the user attempts to achieve 6D localization using the pose related information with respect to multiple bases. For this case, most works \cite{Garcia17,Kanhere21,Kwon21} focus on the position estimation, while the orientation estimation is relatively less investigated. In \cite{Nazari21}, the authors investigated the orientation estimation using \acp{AoA}, assuming the position has been estimated. However, to the best of our knowledge, the joint position and orientation estimation is still missing. At the same time, for the single-BS case, with the increasing depth of the research, the problem of joint position and orientation estimation has been investigated by many works. Enabled by the large bandwidth and antenna arrays with mmWave, in the single-BS case, the estimator exploits the information provided by the multipath to achieve the joint position and orientation estimation. Initial studies on this topic usually place restrictions on the degrees of freedom or focus on planar scenarios. For example, \cite{Shahmansoori18,li2022joint,Mendrzik19,Kakkavas19} studied 2D position and 1D orientation for single-BS case. In these works, the analysis on different aspects, such as estimation method, clock biases, and theoretical bound are more and more profound. In \cite{Wen21,Wen215,Guerra18}, the authors investigated 3D position and 2D orientation under the synchronized condition. Until recently, 3D position and 3D orientation is investigated in \cite{nazari2022mmwave,chen2023modeling}. In \cite{nazari2022mmwave}, the authors studied 3D position and 3D orientation, i.e., 6D localization, in the presence of \ac{LoS} path for the unsynchronized case, where both the estimation method and theoretical bound are analyzed. In \cite{chen2023modeling}, the authors investigated the impact of hardware impairment on the joint estimation of 3D orientation and 3D position under the synchronized condition. However, it can be seen that a unified framework that deals with 6D localization in the absence and presence of \ac{LoS} path %
is still needed. %

In this paper, we reveal a close connection between the \acp{AoA} and \acp{AoD} and 
the well-studied perspective projection model from computer vision \cite{hartley2003multiple}. This connection allows us to solve the  6D localization problem, by adapting state-of-the-art methods from computer vision. More specifically, two problems, namely 6D pose estimation from \acp{AoA} from  multiple base stations and 6D SLAM based on single-BS mmWave communication, are first modeled with the perspective projection model and then solved. The main contributions include:
\begin{itemize}
    \item  We first introduce a projective geometric view for the \ac{AoA}/\ac{AoD} of antenna arrays. We will show that this geometric view links the \ac{AoA}/\ac{AoD} to the well-established perspective projection model. With the help of this model, we derive an explicit expression relating the AOD/AOA to the position of the base station (BS) and the pose of the user equipment (UE).
    \item Two use cases, i.e., 6D pose estimation with \acp{AoA} and SLAM based on mmWave communication, are particularly studied. On the one hand, through the perspective projection modeling, we provide two methods for 6D pose estimation using \acp{AoA}, a closed-form one and an iterative one based on the least squares (LS) principle. On the other hand, for mmWave communication-based SLAM, on the basis of perspective projection modeling, the geometric relation between \ac{AoD}, \ac{AoA}, and scatter points for single BS and single UE scenarios is further modeled with the epipolar model. Based on this modeling, we propose two algorithms for SLAM based on mmWave communication, a closed-form one and an iterative one as well. 
    \item 
    The performance of the proposed algorithm is assessed using Monte Carlo simulations, and the tightness of the results with the \ac{CRB} is evaluated to show that the method is efficient. The proposed method is thoroughly evaluated in a 3D propagation environment, proving its performance at various degrees of surface roughness.

\end{itemize}
The rest of the paper is organized as follows. The projective geometric modeling is presented in Section \ref{sec:background}. The two use cases are investigated in Section III, where the proposed methods as well as the derived theoretical lower bound are given. The Monte Carlo numerical comparison is given in Section IV. Finally, some concluding remarks are given in Section V.

\subsection*{Notations}
We introduce the unit vectors $\mathbf{e}_1=[1\,0\,0]^{\mathrm{T}}$,  $\mathbf{e}_2=[0\,1\,0]^{\mathrm{T}}$, and $\mathbf{e}_3=[0\,0\,1]^{\mathrm{T}}$.  The operator  $\bar{\cdot}$ converts a vector from Cartesian coordinates $\mathbf{x}$ into the homogeneous coordinates, i.e.,  $\bar{\mathbf{x}} = [\mathbf{x}^{\mathrm{T}},1]^{\mathrm{T}}$. A line between two points $\mathbf{x}$ and $\mathbf{y}$ is denoted by  $\overrightarrow{\mathbf{x}\mathbf{y}}$. The operator $\mathbf{x}_{\times}$ generates a skew-symmetric matrix 
\begin{align}\label{crs}
    \mathbf{x}_{\times}=
\begin{bmatrix}
0 &-x_3 & x_2 \\
x_3 &  0  &-x_1 \\
-x_2 & x_1 & 0
\end{bmatrix}.
\end{align}
The operator $\cdot^{\wedge}$ converts a $6\times 1$ vector into a member of $\mathfrak{se}\left(3\right)$ (the Lie algebra of $SE(3)$) by 
\begin{equation}\label{lift}
\left([\mathbf{y}^{\mathrm{T}},\mathbf{x}^{\mathrm{T}}]^{\mathrm{T}}\right)^{\wedge} = 
\begin{bmatrix}
\mathbf{x}_{\times}  & \mathbf{y} \\
\mathbf{0}^{\mathrm{T}} & {0}
\end{bmatrix} \in \mathbb{R}^{4\times4},\quad \mathbf{x},\mathbf{y} \in \mathbb{R}^{3\times 1}.
\end{equation}
We also introduce the shorthand $[x_{\ell \in \{1,2,3\}}]=[x_1, x_2, x_3]$.

\section{System Model and Problem Formulation}
\label{sec:problem}

\begin{figure}
	\centering
	\includegraphics[width=0.49\columnwidth]{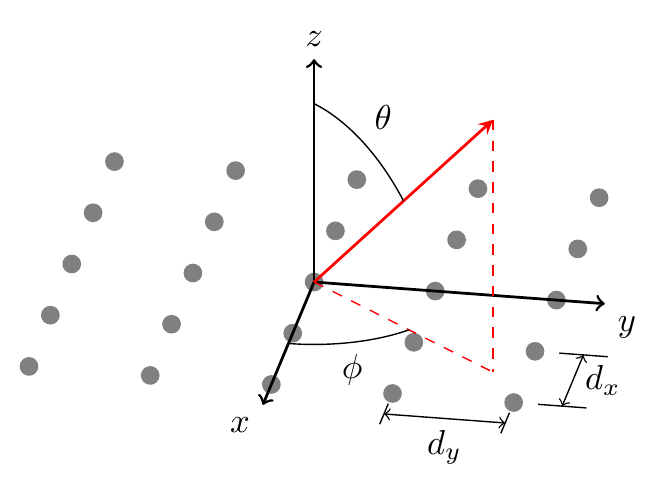}
	\caption{Illustration of an uniform rectangular array along with the azimuth angle $\phi$ and elevation angle $\theta$.}
	\label{fig:URA} \vspace{-5mm}
\end{figure}

\subsection{General System Model}
We consider a scenario with a single mmWave multi-antenna BS with known position $\mathbf{r}_{\mathrm{BS}}$ and a single multi-antenna UE with unknown position  $\mathbf{r}_{\mathrm{UE}}$ and unknown orientation $\mathbf{R}_{\mathrm{UE}} \in SO(3)$. Under downlink transmission, assuming $L\geq 1$ paths between the BS and UE,  the channel for subcarrier $f_n$, represented by $\bm{\mathsf{H}}[n]\in \mathbb{C}^{N_r\times N_t}$, is given by \cite{Zohair18}
\begin{align}
    \label{channel}
     \bm{\mathsf{H}}[n]
    & =  \sum_{\ell\in\mathcal{S}_L}\alpha_{\ell} e^{-j2\pi\tau_{\ell}f_n} \mathbf{a}_{N_r}\left(\bm{\psi}_{\mathrm{R},\ell}\right)\mathbf{a}_{N_t}^{\mathrm{T}}\left(\bm{\psi}_{\mathrm{T},\ell}\right),
\end{align}
where  $\alpha_{\ell}$ is the $\ell^{\text{th}}$ channel gain, $\tau_{\ell}$ the $\ell^{\text{th}}$ propagation delay
and 
\begin{eqnarray}
    \mathcal{S}_L = \left\{
    \begin{array}{cll}
    \{0,\dots,L\} & \quad & \text{for \ac{LoS} case,} \\
    \{1,\dots,L\}  & \quad & \text{for \ac{NLoS} case.}
    \end{array}
    \right. 
\end{eqnarray}
The propagation delays are related to the UE position by
\begin{align}
    \tau_{\ell} & = \begin{cases}
    B + \| \mathbf{r}_{\mathrm{UE}}- \mathbf{r}_{\mathrm{BS}}\|/c & \ell = 0\\
    B+ (\|\mathbf{p}_{\ell} - \mathbf{r}_{\mathrm{BS}}\|+\|\mathbf{p}_{\ell}-\mathbf{r}_{\mathrm{UE}}\|)/c & \ell >0,
    \end{cases}
\end{align}
where $B$ is the UE's unknown clock bias and $\mathbf{p}_{\ell}$ is a location of a scatterer, assuming at most single-bounce reflections.\footnote{Due to the mmWave propagation characteristics, the received power contributed by multiple-bounce reflections is negligible \cite{Shahmansoori18,Hua14}.} 
In addition, $N_r=N_{r,x}N_{r,y}$ is the number antenna of elements in the UE array, comprising $N_{r,x}$ elements along the local x-axis and $N_{r,y}$ along the local y-axis. Similarly,  $N_t=N_{t,x}N_{t,y}$ the number of  antenna elements in BS array. We focus on planar antenna arrays, and, without loss of generality, we consider the uniform rectangular array, as shown in Fig.~\ref{fig:URA}. The angular information of the wavefront, e.g., \ac{AoD}/\ac{AoA}, is conventionally represented by the azimuth and elevation angles $\bm{\psi}=\left[\phi,\theta\right]^{\mathrm{T}}$, which specify the normal vector 
\begin{align}
    \mathbf{n} = \begin{bmatrix}
    \cos\phi\sin\theta \\
    \sin\phi\sin\theta \\
    \cos\theta
    \end{bmatrix}
\end{align}
of the wavefront direction. Then, the steering vector $\mathbf{a}_{N_r}\left(\bm{\psi}_{\mathrm{R},\ell}\right)\in \mathbb{C}^{N_r \times 1}$ is defined as
\begin{align}
    \mathbf{a}_{N_r}\left(\bm{\psi}_{\mathrm{R},\ell}\right) = \bm{\mathfrak{a}}_{1}\left(\bm{\psi}_{\mathrm{R},\ell}\right)\otimes\bm{\mathfrak{a}}_{2}\left(\bm{\psi}_{\mathrm{R},\ell}\right),
\end{align}
with
\begin{align}
     \left[\bm{\mathfrak{a}}_{1}\left(\bm{\psi}\right)\right]_m & =  e^{j\pi m \sin(\theta)\sin(\phi)}, \; m \in \{0,\dots,N_x\}\\
       \left[\bm{\mathfrak{a}}_{2}\left(\bm{\psi}\right)\right]_{m} &=  e^{j\pi m \sin(\theta)\cos(\phi)}, \; m \in \{0,\dots,N_y\},
\end{align}
assuming the antenna spacing is of half-wavelength. 
The \ac{AoD} $\bm{\psi}_{\mathrm{T},\ell}\in\mathbb{R}^{2\times 1}$ and steering vector $\mathbf{a}_{N_t}\left(\bm{\psi}_{\mathrm{T},\ell}\right)$ are defined similarly.

\subsection{Problem 1 --  mmWave MIMO Snapshot SLAM} \label{sec:Problem1}
The objective is to simultaneously estimate the UE pose $\mathbf{r}_{\mathrm{UE}},\mathbf{R}_{\mathrm{UE}} $ as well as the position of scatters $\mathbf{p}_{\ell}$, $\ell\in\left\{1\dots,L\right\}$, based on an estimate of the channels $\bm{\mathsf{H}}[n]$. This objective of mmWave communication-based SLAM is usually achieved through a sequence of multi-step processes \cite{Shahmansoori18,Yu20}: 
\begin{enumerate}
    \item Estimation of the channel matrices $\bm{\mathsf{H}}[n]$ for $n\in\{1,\ldots, N_f\}$ based on the observed pilot signal (this is not the core of this work and will be discussed in Section \ref{sec:results}); 
    \item Estimation of the (effective) parameter vector $\mathbf{z}_{\ell} = \left[\bm{\psi}_{\mathrm{R},\ell},\bm{\psi}_{\mathrm{T},\ell},\tau_{\ell},\alpha_{\ell}\right]^{\mathrm{T}}$ from $\bm{\mathsf{H}}[n]$ for all $N_f$ subcarriers,
    \item Estimation of the UE pose and the position of scatters, i.e., SLAM, based on the parameter vector set $\mathcal{Z} = \{\mathbf{z}_{0},\ldots,\mathbf{z}_{\hat{L}} \}$, where $\hat{L}$ is the detected number of paths.   
\end{enumerate}
Unless stated otherwise, we will assume that \emph{we do not know whether the \ac{LoS} path is present}. 

\subsection{Problem 2 -- \ac{AoA}-only Pose Estimation} \label{sec:Problem2}

\begin{figure}[t]
	\centering
	\includegraphics[width=190pt]{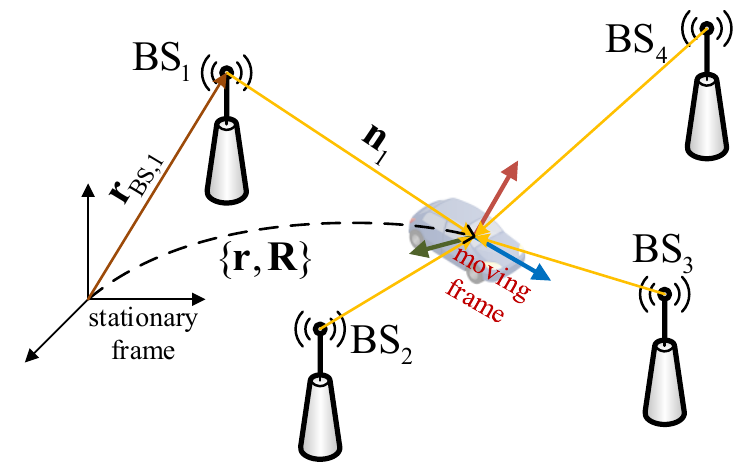}
	\caption{Schematic of 6D pose estimation using the \ac{AoA} with respect to 4 BSs.}\vspace{-5mm}
	\label{Pose} 
\end{figure}

In this problem (as visualized in Fig.~\ref{Pose}), the goal is to estimate the receiver's pose based on the \ac{AoA} with respect to multiple single-antenna BSs at known positions, based on narrowband downlink signals under pure \ac{LoS} propagation. In that case, the channel from BS $i$ to the UE simplifies to $ \bm{{h}}_i  =  \alpha_{i}  \mathbf{a}_{N_r}\left(\bm{\psi}_{\mathrm{R},i}\right)$. 
There are many available algorithms to estimate \ac{AoA}, such as the MUSIC algorithm \cite[Sect. 9.3.2]{van2002optimum}, however, the \ac{AoA}-based pose estimation was only discussed partially, in that either positioning or orientation estimation, e.g., \cite{Jachimczyk17,Nazari21}, is considered. Therefore, the joint estimation of position and orientation, that is, the pose estimation has not been discussed thoroughly.

\section{Background on Computer Vision and Relation to \ac{AoA} and \ac{AoD}} \label{sec:background}
In this section, a short primer on basic results from computer vision is given, followed by the application of these results to expression \acp{AoA} and \acp{AoD} in mmWave MIMO communication systems.

\subsection{Perspective Projection Model}
\begin{figure}[!t]
	\centering
	\includegraphics[width=0.49\columnwidth]{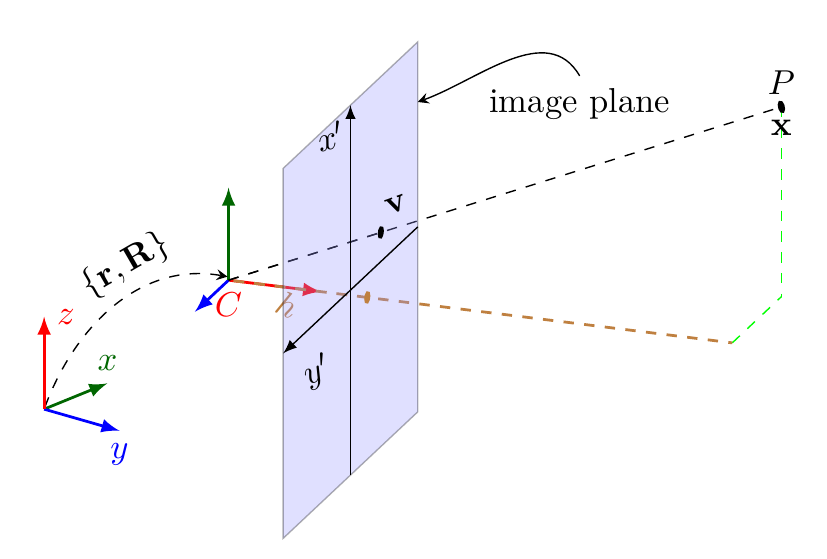}
	\caption{The perspective projection model of a camera with position $\mathbf{r}$ and orientation $\mathbf{R}$, showing the image coordinate $\mathbf{v}$ of a point $P$.}
	\label{fig:pin}\vspace{-5mm}
\end{figure}

Consider an ideal perspective camera and a point $P$ with homogeneous coordinates 
$\bar{\mathbf{x}}_{\mathrm{cam}}\in \mathbb{R}^4$ in the camera frame  and homogeneous coordinates $\bar{\mathbf{x}}\in \mathbb{R}^4$ in the system frame. These coordinates  are linked by the coordinate transformation
    \begin{equation}\label{tranx}
        \bar{\mathbf{x}}_{\mathrm{cam}} = \mathbf{T}\bar{\mathbf{x}},
    \end{equation}
    where the matrix $\mathbf{T} \in SE\left(3\right)$ belongs to the Special Euclidean group $SE\left(3\right)$ and is defined through its inverse
    \begin{equation}\label{TT0}
    \mathbf{T}^{-1} = 
    \begin{bmatrix}
    \mathbf{R} & \mathbf{r} \\
    \mathbf{0}_{1\times3} & 1 
    \end{bmatrix}, %
    \end{equation}
in which $\{\mathbf{r},\mathbf{R}\}$ represent the pose of the camera, comprising a displacement vector $\mathbf{r} \in \mathbb{R}^3$ and a rotation matrix $\mathbf{R} \in SO(3)$, depicted in Fig.~\ref{fig:pin}. 

The image coordinates 
$\mathbf{v} \in \mathbb{R}^2$ of $P$ (see Fig.~\ref{fig:pin}) are 
given by the so-called perspective projection model as \cite[Sect. 6.1]{hartley2003multiple}
    \begin{equation}\label{ppm}
        s\bar{\mathbf{v}} = \mathbf{P}\bar{\mathbf{x}}_{\mathrm{cam}},
    \end{equation}
    where %
    $\bar{\mathbf{v}}$ are the homogeneous coordinate of the image point $\mathbf{v}$, $s$ is a scale factor ($s = 1/\mathbf{e}_3^{\mathrm{T}}{\mathbf{x}_{\mathrm{cam}}}$), and $\mathbf{P} \in \mathbb{R}^{3 \times 4}$ is a projection matrix, given by  %
    \begin{align}
        \mathbf{P}=\begin{bmatrix}
    h, 0, 0 \\
    0, h, 0 \\
    0, 0, 1
    \end{bmatrix} [\mathbf{I}_{3\times 3}\; \mathbf{0}_{3\times 1}], 
    \end{align}
    in which $h$ the camera focal length (i.e., the distance of the image plane).

    Taking into account \cref{ppm,tranx,TT0} and keeping only the first two elements of $\bar{\mathbf{v}}$, we have a concise relation between 
the 2D image coordinates $\mathbf{v}$ of a 3D point $P$ and its 3D coordinates $\mathbf{x}$ in the system frame: 
    \begin{equation}\label{prjp0}
    {\mathbf{v}}= \frac{{\mathbf{K}}\mathbf{T}\bar{\mathbf{x}}}{\mathbf{e}_3^{\mathrm{T}}\mathbf{T}\bar{\mathbf{x}}} \in \mathbb{R}^{2\times 1},
    \end{equation}
    where 
    \begin{equation}\label{KK}
    {\mathbf{K}} = 
    \begin{bmatrix}
    h, 0, 0, 0 \\
    0, h, 0, 0
    \end{bmatrix}
     \in \mathbb{R}^{2\times 4}.
    \end{equation}
\subsection{Epipolar Model}\label{sec:EpipolarModel}

\begin{figure}[!t]
	\centering
	\includegraphics[width=0.355\textwidth]{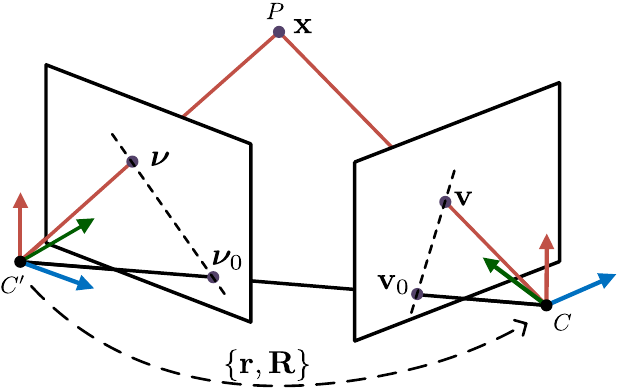}
	\caption{The epipolar modeling, involving the image coordinates with respect to two cameras ($C,C'$) of a common point $P$ with system coordinates $\mathbf{x}$.  The epipoles are $\mathbf{v}_0$ and $\bm{\nu}_0$. Two epipolar lines are shown in dashed. }\vspace{-5mm}
	\label{epipolar}
\end{figure}
Consider now two cameras, denoted by 
 $C$ and $C'$ and a point $P$. 
  We now denote by $\{\mathbf{r},\mathbf{R}\}$  the relative pose of camera $C$ to camera $C'$ (see Fig.~\ref{epipolar}).  The point $P$ leads to 
 two image coordinates, say $\bm{\nu}$ and $\mathbf{v}$, which are related to one another, via the relative pose $\{\mathbf{r},\mathbf{R}\}$. In this section, we will describe this relation.

 The two camera centers can be connected by a line. This line  is called the baseline in computer vision, and it intersects the image planes at two points, i.e., $\mathbf{v}_0$ and $\bm{\nu}_0$, which are called the \emph{epipoles}. The line joining a image point and the epipole in each image plane is called an \emph{epipolar line}, e.g., the line $\overrightarrow{\mathbf{v}_0\mathbf{v}}$.
It follows that for each point $\mathbf{v}$ in one image, there exists a corresponding epipolar line $\overrightarrow{\bm{\nu}_0\bm{\nu}}$ in the other image. Any point $\bm{\nu}$ in the second image matching the point $\mathbf{v}$ must lie on the epipolar line $\overrightarrow{\bm{\nu}_0\bm{\nu}}$. The above relation between these image points is characterized by the epipolar model in computer vision, which is \cite[Sect. 9.6]{hartley2003multiple}
\begin{equation}\label{epiE}
    \bar{\bm{\nu}}^{\mathrm{T}}\mathbf{E}\bar{\mathbf{v}} = 0,
\end{equation}
where 
\begin{equation}\label{ErR}
    \mathbf{E} = \mathbf{r}_{\times}\mathbf{R} \in \mathbb{R}^{3 \times 3} 
\end{equation}
is the so-called \emph{essential matrix}. The essential matrix is used in computer vision to determine relative poses between two cameras, based on matched image points.

\subsection{A Projective Geometric View of the \ac{AoD}/\ac{AoA}}

\begin{figure}[t]
	\centering
	\includegraphics[width=0.49\columnwidth]{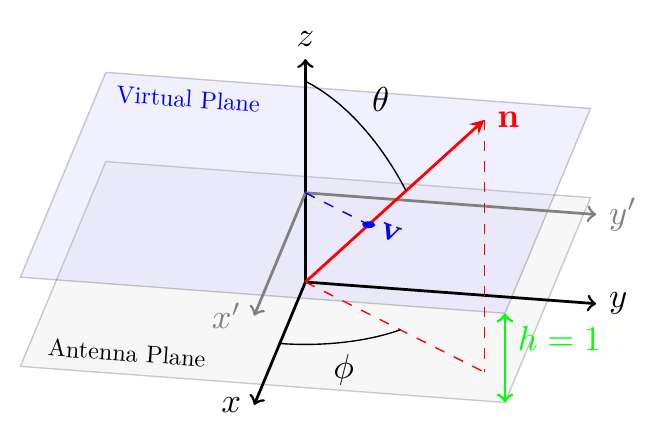}
	\caption{Illustration of the virtual plane of a uniform rectangular array.}
	\label{fig:URA_vp} \vspace{-5mm}
\end{figure}

We are now ready to relate the perspective project model and the epipolar model to \acp{AoA} and \acp{AoD} in mmWave MIMO systems. Due to the antenna reciprocity, we focus on the \ac{AoA} at the receiver end in the following, and similar conclusion can be inferred for the \ac{AoD} at transmitter end. 

We recall that the
angular information of the wavefront, e.g., \ac{AoD}/\ac{AoA}, is conventionally represented by the azimuth and elevation angles $(\phi,\theta)$, which specify the normal vector 
\begin{equation}\label{unitv}
    \mathbf{n} = \begin{bmatrix}
    \cos\phi\sin\theta \\
    \sin\phi\sin\theta \\
    \cos\theta
    \end{bmatrix}
\end{equation}
of the wavefront direction.
Consider now the physical antenna plane (say, the XY plane in the array frame) and a a virtual plane parallel to the XY plane at $z=1$, as shown in Fig.~\ref{fig:URA_vp}. 
It can be readily seen that  each wavefront direction $\mathbf{n}\in \mathbb{R}^3$ can be bijectively mapped to a 2D point on the plane, i.e., the intersection $\mathbf{v}\in\mathbb{R}^{2\times1}$ of the virtual plane and the line of the wavefront direction through the origin, with (in homogeneous coordinates)
\begin{equation}\label{prjv}
    \bar{\mathbf{v}} = \frac{\mathbf{n}}{(\mathbf{e}^{\mathrm{T}}_3\mathbf{n})} = 
    \begin{bmatrix}
    \cos\phi\tan\theta \\
    \sin\phi\tan\theta \\
    1
    \end{bmatrix}.
\end{equation}
The line $\overrightarrow{\mathbf{r}_{\mathrm{BS}}\mathbf{r}_{\mathrm{UE}}}$ specifies the wavefront direction at the UE receiver end of the direct path. 
By analogy with a camera, if the BS is seen as the object point, the UE antenna center as the camera center, after introducing the virtual plane, 
$\mathbf{v}$ can be seen as the ``projected" point on the virtual plane at the UE of the point $\mathbf{r}_{\mathrm{BS}}$ along the line $\overrightarrow{\mathbf{r}_{\mathrm{BS}}\mathbf{r}_{\mathrm{UE}}}$, complying with the perspective projection model from Fig.~\ref{fig:pin}.  Consequently, if we further assume that the orientation of the receiver in the system frame is described by $\mathbf{R}_{\mathrm{UE}} \in SO(3)$,
then, following \eqref{prjp0}, the coordinates $\mathbf{v}$ are given by
\begin{equation}\label{prjp}
{\mathbf{v}}= \frac{{\mathbf{K}}\mathbf{T}_{\mathrm{UE}}\bar{\mathbf{r}}_{\mathrm{BS}}}{\mathbf{e}_3^{\mathrm{T}}\mathbf{T}_{\mathrm{UE}}\bar{\mathbf{r}}_{\mathrm{BS}}} \in \mathbb{R}^{2\times 1},
\end{equation}
where the matrix $\mathbf{T}_{\mathrm{UE}} \in SE\left(3\right)$ is defined through its inverse
\begin{equation}\label{TT}
\mathbf{T}^{-1}_{\mathrm{UE}} = 
\begin{bmatrix}
\mathbf{R}_{\mathrm{UE}} & \mathbf{r}_{\mathrm{UE}} \\
\mathbf{0}_{1\times3} & 1 
\end{bmatrix}, %
\end{equation}
and ${\mathbf{K}}$ was defined in \eqref{KK}, with $h=1$. 
Hence, based on \cref{unitv,prjv,prjp}, %
we have related the angular information (represented by either $\mathbf{v}$, $\mathbf{n}$ or $(\theta,\phi)$) to the BS position $\mathbf{r}_{\mathrm{BS}}$, and UE pose $\mathbf{T}_{\mathrm{UE}}$ with the perspective projection model. For the \ac{AoD}, we have a similar relation, i.e.,
\begin{equation}\label{prjpAOD}
{\boldsymbol{\nu}}= \frac{{\mathbf{K}}\mathbf{T}_{\mathrm{BS}}\bar{\mathbf{r}}_{\mathrm{UE}}}{\mathbf{e}_3^{\mathrm{T}}\mathbf{T}_{\mathrm{BS}}\bar{\mathbf{r}}_{\mathrm{UE}}} \in \mathbb{R}^{2\times 1},
\end{equation}
where $\mathbf{T}_{\mathrm{BS}}$ specifies the pose of the BS.

\section{Projective Geometry Solutions to mmWave MIMO Pose Estimation}
Based on the relations \cref{prjp,prjpAOD}, we can now reformulate the problems from Section \ref{sec:Problem1} and Section \ref{sec:Problem2} in terms of a perspective projection model, in order to apply methods from computer vision. We start with Problem 2, as it is less complex.

\subsection{Solution to Problem 2 -- \ac{AoA}-only Pose Estimation} \label{sec:SolutionProblem2}
The perspective projection model provides us with a new mathematical description to simplify the problem from Fig.~\ref{Pose}. After obtaining virtual points from \acp{AoA}, 
the problem is converted into estimating the UE pose from 3D to 2D point correspondences $(\mathbf{r}_{\mathrm{BS},i},\mathbf{v}_{i})$, $i=1,\ldots, I$ in analogy to the problem of estimating the camera's pose from 3D landmark to 2D image correspondences.
We denote by $\tilde{\mathbf{V}}=\left[\tilde{\mathbf{v}}_1,\dots,\tilde{\mathbf{v}}_{I}\right]$ the observation matrix containing the observed virtual points $\tilde{\mathbf{v}}_i$ converted from \ac{AoA} $(\phi_i,\theta_i)$ according to \eqref{prjv}, for $i\in\{1,\dots,I\}$,
We also introduce $\mathbf{V}(\mathbf{T}_{\mathrm{UE}}) = \left[\mathbf{v}_1,\dots,\mathbf{v}_{I}\right]$ as a function of $\mathbf{T}_{\mathrm{UE}}$ with $\mathbf{v}_i$ modeled by \eqref{prjp}. With these formulations, standard %
computer vision methods can be applied, 
including closed-form solutions such as the Perspective-n-Point (PnP) algorithm \cite[12.2]{forstner2016photogrammetric}, as well as the iterative method given by
\begin{subequations}
\label{mle}
\begin{align}
    \text{minimize} & \quad \left\|\tilde{\mathbf{V}}-\mathbf{V}(\mathbf{T}_{\mathrm{UE}} )\right \|^2_F \\
    \text{s.t.} &\quad  \mathbf{T}_{\mathrm{UE}} \in SE(3)
\end{align}
\end{subequations}
in order to solve for $\mathbf{T}_{\mathrm{UE}}$. 
The problem \eqref{mle}
can be solved by off-the-shelf algorithm toolboxes for cameras, and the closed-form algorithms, such as the PnP algorithm, can be used for initialization.

Finally, we note that since that the pose has six degrees of freedom and that each correspondence generates two constraints, at least $I=3$  bases are needed to estimate the pose \cite[Sect. 7.3]{hartley2003multiple}.

\subsection{Solution to Problem 1 --  mmWave MIMO Snapshot SLAM} \label{solutionProblem1}

\begin{figure}%
	\centering
    \includegraphics[width=0.49\textwidth]{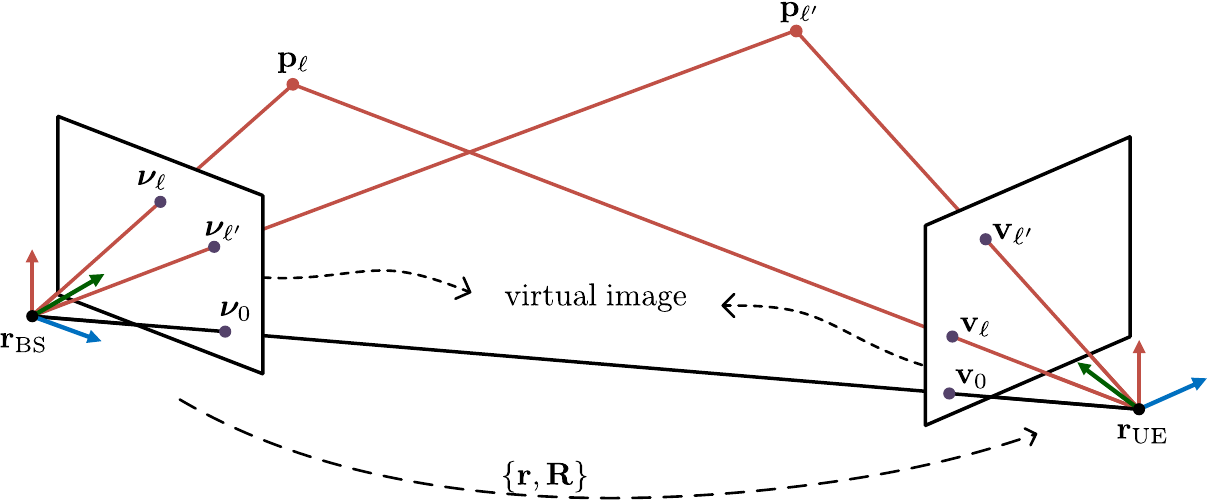}
	\caption{Illustration of 3D MIMO channel in the epipolar modeling.}\vspace{-5mm}
	\label{Epipolar} 
\end{figure}

Moving on to Problem 1, we focus on step 3 from Section \ref{sec:Problem1}, considering that estimates of the AOAs and AODs are given. To simplify the analysis, we will set the BS pose so that $\mathbf{R}_{\mathrm{BS}}=\mathbf{I}_3$ and $\mathbf{r}_{\mathrm{BS}}=\mathbf{0}_{3 \times 1}$, i.e., set the system frame to align with the BS antenna, then the UE pose is given by the relative pose to the BS. 
We apply the perspective projection model to both the UE and BS antenna arrays, so that each scatter point makes a virtual point on each virtual plane, as shown in Fig.~\ref{Epipolar}. It can be seen that (i) the \ac{LoS} path is represented by the baseline, which intersects the virtual planes at $\mathbf{v}_0$ (at UE) and $\bm{\nu}_0$ (at BS); (ii) each scatter point $\mathbf{p}_{\ell}$ is ``projected" in two virtual points, at $\mathbf{v}_{\ell}$  (at the UE), and $\bm{\nu}_{\ell}$ (at the BS). 
As a result, the geometric relation between the transmitter, receiver and scatter points can be described by the epipolar model from Section \ref{sec:EpipolarModel}: %
\begin{equation}\label{epipath}
    \bar{\bm{\nu}}^{\mathrm{T}}_{\ell}\mathbf{E}\bar{\mathbf{v}}_{\ell} = 0, \quad \mathrm{for} \; \ell\in\mathcal{S}_L,
\end{equation}
where the essential matrix $\mathbf{E} = (\mathbf{r}_{\mathrm{UE}})_{\times}\mathbf{R}_{\mathrm{UE}}$ fully describes the UE pose. 

We now proceed to solve the SLAM problem, relying to existing computer vision methods and applying modifications where needed. Note that we ignore effects such as angle estimation errors and channel gain estimates, in order to obtain a closed-form SLAM methods,  summarized in Alg.~\ref{alg:cap}. 
These effects  can be later included in suitable maximum-likelihood methods, initialized by our approach. We also ignore the effect that  the paths in \eqref{channel} may not all be resolvable. This effect will be considered in the numerical results in Section \ref{sec:results}.

We start from estimates of the \ac{AoA} $\tilde{\bm{\psi}}_{\mathrm{R},\ell}$, \ac{AoD}  $\tilde{\bm{\psi}}_{\mathrm{T},\ell}$, and delays $\tilde{\tau}_{\ell}$, for $\ell\in\mathcal{S}_L$. Here $\tilde{\cdot}$ is used to denote observations (inputs), while $\hat{\cdot}$ is used to denote estimates (outputs).

\begin{algorithm}
\caption{Summary of the Closed-Form Algorithm}\label{alg:cap}
\begin{algorithmic}
\Require $\tilde{\bm{\psi}}_{\mathrm{R},\ell}$ $\tilde{\bm{\psi}}_{\mathrm{T},\ell}$, $\tilde{\tau}_{\ell}$, for $\ell\in\mathcal{S}_L$
\State 
Obtain $\tilde{\bm{\nu}}_{\ell}$, and $\tilde{\mathbf{v}}_{\ell}$ based on \eqref{prjv}
\State Estimate essential matrix $\hat{\mathbf{E}}$ based on $\tilde{\bm{\nu}}_{\ell}$, and $\tilde{\mathbf{v}}_{\ell}$
\State Estimate $\hat{\mathbf{R}}_{\mathrm{UE}}$, and $\hat{\mathbf{n}}_r$ based on $\hat{\mathbf{E}}$
\State Triangulation for $\breve{\mathbf{p}}_{\ell}$ based on $\hat{\mathbf{R}}_{\mathrm{UE}}$, $\hat{\mathbf{n}}_{\mathbf{r}}$, $\tilde{\mathbf{v}}_{\ell}$ and $\tilde{\bm{\nu}}_{\ell}$
\State Metric reconstruction of $\hat{s}$ \eqref{sfactor}, and recover  $\hat{\mathbf{p}}_{\ell}$ and $\hat{\mathbf{r}}_{\mathrm{UE}}$,
\end{algorithmic}
\end{algorithm}

\subsubsection{Phase 1: Estimation of the virtual points and the essential matrix from the \acp{AoD} and \acp{AoA}}
We first convert \acp{AoD} and \acp{AoA} estimates into the virtual points, i.e., from  $\tilde{\bm{\psi}}_{\mathrm{R},\ell}$ and $\tilde{\bm{\psi}}_{\mathrm{T},\ell}$ to $\mathbf{v}_{\ell}$ and $\bm{\nu}_{\ell}$ for $\ell\in\mathcal{S}_L$, based on \eqref{prjv}. 
    Then, based on multiple pairs of $\mathbf{v}_{\ell}$ and $\bm{\nu}_{\ell}$, the essential matrix $\mathbf{E}$ can be estimated by using, for example, the algorithm given in  \cite{nister2004efficient,kukelova2008polynomial}.%

\subsubsection{Phase 2: Computation of the relative UE pose from the estimate $\hat{\mathbf{E}}$}
This step can be implemented by applying 
    the SVD-based algorithm from \cite[Sects. 9.6.2 and 9.6.3]{hartley2003multiple} to $\hat{\mathbf{E}}$ to estimate the orientation $\mathbf{R}_{\mathrm{UE}}$ and the normal vector of the position $\mathbf{n}_{\mathbf{r}} = \mathbf{r}_{\mathrm{UE}}/\|\mathbf{r}_{\mathrm{UE}}\|$.
    Note that in this step, the UE position can only be estimated up to scale, therefore, the position is only estimated up to the unit vector $\mathbf{n}_{\mathbf{r}}$.

\subsubsection{Phase 3: Triangulation for computing the 3D scatter points}
The scatter positions are estimated (again up to a scale) and denoted by $\breve{\mathbf{p}}_{\ell}, \ell\in\{1,\dots,L\}$,  based on $\hat{\mathbf{R}}_{\mathrm{UE}}$, $\hat{\mathbf{n}}_{\mathbf{r}}$, $\tilde{\mathbf{v}}_{\ell}$ and $\tilde{\bm{\nu}}_{\ell}$ by using, for example, the homogeneous method \cite[Sect. 12.2]{hartley2003multiple}. %

\subsubsection{Phase 4: Scale recovery}
In order to  recover the scale factor so that the relative UE pose and scatter positions can be fully recovered, computer vision methods require knowledge of the overall scale of the scene. As this is not possible in mmWave MIMO pose estimation, we instead rely on the estimated propagation delays. %
We introduce the vector of estimated delays, $\tilde{\bm{\tau}}$ with $[\tilde{\bm{\tau}}]_{\ell}=\tilde{\tau}_{\ell}$, $\ell \in \mathcal{S}_L$, and a corresponding vector of scaled path lengths $[\breve{\mathbf{d}}]_{\ell}=\breve{d}_{\ell}$, $\ell \in \mathcal{S}_L$, and $\breve{d}_{\ell} = \|\breve{\mathbf{p}}_{\ell}\|+\|\breve{\mathbf{p}}_{\ell}-\hat{\mathbf{n}}_{\mathbf{r}}\|$. Under the correct scaling $s$, we recall that  $\tilde{\tau}_{\ell}=s \breve{d}_{\ell}/c+B$. To get rid of the clock bias, we introduce differential measurements  $\mathbf{D}\tilde{\bm{\tau}}$ and $\mathbf{D}\breve{\mathbf{d}}$, where $\mathbf{D}$   is a $|\mathcal{S}_L| \times |\mathcal{S}_L|$ binary matrix. This matrix is constructed as $\mathbf{D} = \mathbf{I}_{|\mathcal{S}_L|}-\mathbf{D}'$, where $\mathbf{D}'$ has exactly one `1' on each each row, off the main diagonal.\footnote{In the special case where UE and BS are synchronized, $B$ is known to be $0$, so that $\mathbf{D} = \mathbf{I}_{|\mathcal{S}_L|}$ can be used.} Then, $\mathbf{D}\tilde{\bm{\tau}}=s\mathbf{D}\breve{\mathbf{d}}/c$, so that the scaling is found as 
    \begin{equation}\label{sfactor}
        \hat{s} = \frac{c}{\big|\mathcal{S}_L\big|}\sum_{\ell\in\mathcal{S}_L}\frac{[\mathbf{D}_s\tilde{\bm{\tau}}]_{\ell}}{[\mathbf{D}_s\breve{\mathbf{d}}]_{\ell}}.
    \end{equation}
    Finally, the positions of the receiver and scatter point are recovered by $\hat{\mathbf{r}}_{\mathrm{UE}} = \hat{s}\hat{\mathbf{n}}_{\mathbf{r}}$, and $\hat{\mathbf{p}}_{\ell} = \hat{s}\breve{\mathbf{p}}_{\ell}$, respectively.

\subsection{Problem 1 - Discussion, Variations, and Refinement}

\subsubsection{Identifiability without \ac{LoS} knowledge}
The epipolar model also gives the minimum number of distinctive scatter points required for SLAM using the estimates of \ac{AoD}, \ac{AoA}, and propagation delay: %
at least five distinctive pairs of virtual point correspondences are required to estimate the essential matrix \cite[Sect. 13.3.2]{forstner2016photogrammetric} by, for example, using the five point algorithm (see Appendix \ref{sect:append_5pt} for a brief introduction and refer to \cite{nister2004efficient,kukelova2008polynomial} for more details). Therefore, in order to achieve the SLAM problem of the 6D UE pose and 3D scatter positions estimation using the estimates of \ac{AoD}, \ac{AoA}, and propagation delay, the estimator requires four and five distinctive scatter points for the \ac{LoS} and \ac{NLoS} scenarios, respectively. %

\subsubsection{Identifiability with \ac{LoS} knowledge}
If we know that the \ac{LoS} path is present, the methods can be further improved (refer to Appendix \ref{sect:append_los}) and the number of scatter points can be reduced to one, in line with the literature \cite{nazari2022mmwave}. {Note that in this case, the element within $\breve{\mathbf{d}}$ in \eqref{sfactor} corresponding to the \ac{LoS} path is a constant 1, since the scaled path lengths are normalized so that the scaled \ac{LoS} distance is norm-1.}

\subsubsection{Iterative Refinement}
In addition to the above closed-form solution consisting of sequential steps, we propose a direct estimation based on the weighted \ac{LS} principle, which is given by
\begin{align}\label{Obj}
\hat{\mathbf{\Theta}}=&\arg\min_{\mathbf{\Theta}}\mathcal{L}(\mathbf{y},\mathbf{\Theta}), \nonumber \\
& s.t. \; \mathbf{T}_{\mathrm{UE}} \in SE(3)
\end{align}
where the parameter set is $\mathbf{\Theta}=\{\mathbf{T}_{\mathrm{UE}},\mathbf{P}\}$, $\mathbf{P} = \left[\mathbf{p}_1,\dots,\mathbf{p}_L\right]$, and the observation vector $\mathbf{y} = \mathrm{vec}(\mathbf{y}_{\ell \in \mathcal{S}_L})$ with $\mathbf{y}_{\ell} = \left[\tilde{\bm{\nu}}^{\mathrm{T}}_{\ell},\tilde{\mathbf{v}}^{\mathrm{T}}_{\ell},\tilde{\tau}_{\ell}\right]^{\mathrm{T}}$. %
In \eqref{Obj}, the objective function $\mathcal{L}(\mathbf{y},\mathbf{\Theta})$ is given by 
\begin{align}
& \mathcal{L}(\mathbf{y},\mathbf{\Theta}) =
\\ 
& \sum_{\ell\in\mathcal{S}_L}w_{1,\ell}\|\tilde{\bm{\nu}}_{\ell}-\bm{\nu}_{\ell}\|^2 +%
w_{2,\ell}\|\tilde{\mathbf{v}}_{\ell}-\mathbf{v}_{\ell}\|^2 + %
w_{3,\ell}\left[\mathbf{D}_s(\tilde{\bm{\tau}} - \bm{\tau})\right]_{\ell}^2,\nonumber
\end{align}
where $w_i$ for $i\in\{1,2,3\}$ is the weight factor and should properly reflect the precision of estimates $\tilde{\bm{\psi}}_{\mathrm{R},\ell}$, $\tilde{\bm{\psi}}_{\mathrm{T},\ell}$ and $\tilde{\tau}_{\ell}$. The constraint in (\ref{Obj}) implies $\mathbf{R}_{\mathrm{UE}}^{\mathrm{T}}\mathbf{R}_{\mathrm{UE}} = \mathbf{I}
, \det\left(\mathbf{R}_{\mathrm{UE}}\right) = +1$, and considered that $SE\left(3\right)$ is a manifold, the optimization of \eqref{Obj} can be solve by the Gauss-Newton method on the corresponding manifold \cite{absil2009optimization} to convert the above optimization problem into an unconstrained optimization problem on the manifold, which is obtained with an iterative procedure. At each iteration, the update step is \cite[Eq. 7.196]{barfoot_2017}
\begin{align}\label{upk}
\mathbf{T}^{t+1}_{\mathrm{UE}} & = \exp\left(\left(\kappa^t_{\mathbf{T}}\mathbf{\Delta}^t_{\mathbf{T}}\right)^{\wedge}\right) \mathbf{T}^{t}_{\mathrm{UE}}, \\
\mathbf{p}_{\ell}^{t+1} & = \mathbf{p}_{\ell}^{t} + \kappa^t_{\mathbf{p}_{\ell}}\mathbf{\Delta}^t_{\mathbf{p}_{\ell}} \quad \ell \in \{1,\dots,L\}
\end{align}
where $\kappa^t_{\mathbf{T}}>0$ and $\kappa^t_{\mathbf{p}_{\ell}}>0$ control the
incremental step size for $\mathbf{T}_{\mathrm{UE}}$ and $\mathbf{p}_{\ell}$, respectively. 
In \eqref{upk}, the update direction is calculated by
\begin{equation}\label{gnm}
\left[\mathbf{\Delta}_{\mathbf{T}},\mathbf{\Delta}_{\mathbf{p}_1},\dots,\mathbf{\Delta}_{\mathbf{p}_{L}}\right]^{\mathrm{T}}=\left(\nabla_{\mathbf{\Theta}}\bm{\mu}\right)^{\dagger}\left(\mathbf{y}-\bm{\mu}\right)
\end{equation}
where $(\cdot)^{\dagger}$ is the weighted pseudoinverse defined by \cite[Eq. 7.32]{siciliano2008springer}
\begin{equation}
    (\mathbf{J})^{\dagger} = (\mathbf{J}^{\mathrm{T}}\mathbf{W}\mathbf{J})^{-1}\mathbf{J}^{\mathrm{T}}\mathbf{W},
\end{equation}
$\mathbf{W} = \mathrm{diag}(\mathrm{vec}(\mathbf{w}_{\ell\in\mathcal{S}_L}))$, $\mathbf{w}_{\ell} = [w_{1,\ell},w_{1,\ell},w_{2,\ell},w_{2,\ell},w_{3,\ell}]^{\mathrm{T}}$ $\bm{\mu} = \mathrm{vec}(\bm{\mu}_{\ell \in \mathcal{S}_L}) \in\mathbb{R}^{5|\mathcal{S}_L|\times1}$ with
$\bm{\mu}_{\ell} = \left[\bm{\nu}_{\ell}^{\mathrm{T}}, \mathbf{v}_{\ell}^{\mathrm{T}},\left[\mathbf{D}_s\bm{\tau}\right]_{\ell} \right]^{\mathrm{T}}$,
$\nabla_{\bm{\Theta}}\bm{\mu} = [\nabla_{\bm{\Theta}}^{\mathrm{T}}{\bm{\mu}}_{\ell \in \mathcal{S}_L}]^{\mathrm{T}}\in\mathbb{R}^{5|\mathcal{S}_L|\times\left(6+3L\right)}$ with $\nabla_{\bm{\Theta}}{\bm{\mu}}_{\ell} = \left[\nabla_{\bm{\Theta}}^{\mathrm{T}}\bm{\nu}_{\ell},\nabla_{\bm{\Theta}}^{\mathrm{T}}\mathbf{v}_{\ell}, \left[\mathbf{D}_s\nabla_{\bm{\Theta}}^{\mathrm{T}}\bm{\tau}\right]_{\ell}\right]^{\mathrm{T}}$ the gradient of $\bm{\mu}_{\ell}$ with respect to $\bm{\Theta}$. The
 involving derivatives are given in the Appendix \ref{app:gradients}. 
Further, the initialization of the Gauss-Newton method can be achieved with the closed-form solution given previously.

\subsubsection{Computation Complexity}
For the closed-form algorithm with a set of 5 paths, 
in phase 1, the calculation of virtual points has a complexity of $\mathcal{O}(2\times7)$. In addition, the computation of the essential matrix is dominated by the SVD for \eqref{fpa} and by the matrix inversion and SVD for calculating the solution to the 10 third-order polynominal equations. These two SVDs and the matrix inversion have a complexity of  $\mathcal{O}(5\times9^2)$, $\mathcal{O}(10^3)$, and $\mathcal{O}(10^3)$, respectively. In phase 2, the SVD-based algorithm is dominated by the computation of a SVD and a matrix multiplication, which have a complexity of $\mathcal{O}(3^3)$, and $\mathcal{O}(3^3)$, respectively. In phase 3, the computation of the homogeneous method for each path is dominated by the SVD of a $4\times 4$ matrix, which leads a complexity of $\mathcal{O}(5\times4^3)$. In phase 4, the calculation of  the scale factor has a complexity of $\mathcal{O}(21)$.
Hence, for 5 paths, the complexity is dominated by the estimation of the essential matrix in phase 1. 

To account for more than 5 paths, we rely on RANSAC. In this strategy \cite{nister2004efficient}, a number of random samples containing five correspondences each are first taken. Then the five-point algorithm is applied to each sample and the estimated essential matrix is scored with the Sampson distance \cite[Sect. 11.4.3]{hartley2003multiple}. The estimate with the best score is chosen as the final estimate. 

Finally, for the iterative refinement, the complexity is dominated by the computation of the pseudoinverse, which has a complexity of $\mathcal{O}((6+3|\mathcal{S}_L|)(5|\mathcal{S}_L|)^2)$ per iteration.

\section{Numerical Results} \label{sec:results}
In this section, we numerically analyze the performance of the proposed estimators in the two problems from Section \ref{sec:problem}. 

\subsection{Error metrics}
The error vector measuring the residual error of the estimate is defined as $\bm{\epsilon}(\mathring{\mathbf{\Theta}},\mathbf{\Theta}) = \left[\bm{\epsilon}_{\mathbf{r}}^{\mathrm{T}},\bm{\epsilon}_{\mathbf{u}}^{\mathrm{T}},\bm{\epsilon}_{\mathbf{p}}^{\mathrm{T}}\right]^{\mathrm{T}}$ with $\bm{\epsilon}_{\mathbf{r}}=\mathring{\mathbf{r}}_{\mathrm{UE}}-\mathbf{r}_{\mathrm{UE}}$, $\bm{\epsilon}_{\mathbf{u}}=\log(\mathbf{R}_{\mathrm{UE}}\mathring{\mathbf{R}}^{-1}_{\mathrm{UE}})^{\vee}$, and $\bm{\epsilon}_{\mathbf{p}}=\mathrm{vec}(\mathring{\mathbf{P}}-\mathbf{P})$, where $\bm{\epsilon}_{\mathbf{u}}$ measures of orientation error vector between $\mathbf{R}_{\mathrm{UE}}$ and $\mathring{\mathbf{R}}_{\mathrm{UE}}$ \cite{Shen19}.
The performance of the estimator is measured by the  \acp{RMSE},  which are defined as
\begin{align}
\text{RMSE}_{\mathbf{r}}& = \sqrt{\mathbb{E}\left\{\bm{\epsilon}_{\mathbf{r}}^{\mathrm{T}}\bm{\epsilon}_{\mathbf{r}}\right\}}  \\
\text{RMSE}_{\mathbf{u}} &= \sqrt{\mathbb{E}\left\{\bm{\epsilon}_{\mathbf{u}}^{\mathrm{T}}\bm{\epsilon}_{\mathbf{u}}\right\}}  \\
\text{RMSE}_{\mathbf{p}} &= \sqrt{\mathbb{E}\left\{\bm{\epsilon}_{\mathbf{p}}^{\mathrm{T}}\bm{\epsilon}_{\mathbf{p}}\right\}},
\end{align}
where $\mathbb{E}\{ \cdot\}$ is the expectation operator. 
\subsection{Problem 2 -- \ac{AoA}-only Pose Estimation}
In this section, we evaluate the method proposed in Section \ref{sec:SolutionProblem2}, for determining the pose of a UE based on signal from several single-antenna BSs. 
\subsubsection{Scenario}
In evaluating the problem of using the \ac{AoA} from multiple BSs for pose estimation, we consider a scenario with $4$ single-antenna BSs at $\mathbf{r}_{\mathrm{BS},1} = \left[-24,-20,8.5\right]^{\mathrm{T}}$~m, $\mathbf{r}_{\mathrm{BS},2} = \left[25,-25,9\right]^{\mathrm{T}}$~m, $\mathbf{r}_{\mathrm{BS},3} = \left[-22, 20,8\right]^{\mathrm{T}}$~m, and $\mathbf{r}_{\mathrm{BS},4} = \left[23,25, 10\right]^{\mathrm{T}}$~m. The UE is equipped with a uniform rectangular array of $10\times 10$ elements at half-wavelength spacing. The carrier frequency is 28 GHz, and $i^{\text{th}}$ BS is assigned with a baseband transmit signal of $s_i(t) = e^{j2\pi f_{s,i} t}$ with $f_{s,i} = 100 + 50i$~Hz. The \ac{SNR} is defined as $\text{SNR}=|s_i|^2/\sigma^2_s$, where $\sigma^2_s$ is the variance of the additive Gaussian noise. At the receiver side, the MUSIC algorithm \cite[Sect. 9.3.2]{van2002optimum} is used to estimate \ac{AoA}, and then the association of the AOA estimate and corresponding BS is achieved by MVDR beamforming \cite[Sect. 6.2.1]{van2002optimum} to identify the transmit signal frequency. Finally, the P3P algorithm \cite{Gao03} is used to estimate the pose, using the virtual points converted from the AOA estimates.
\begin{figure}[!tb]
	\centering
	\includegraphics[width=0.495\textwidth]{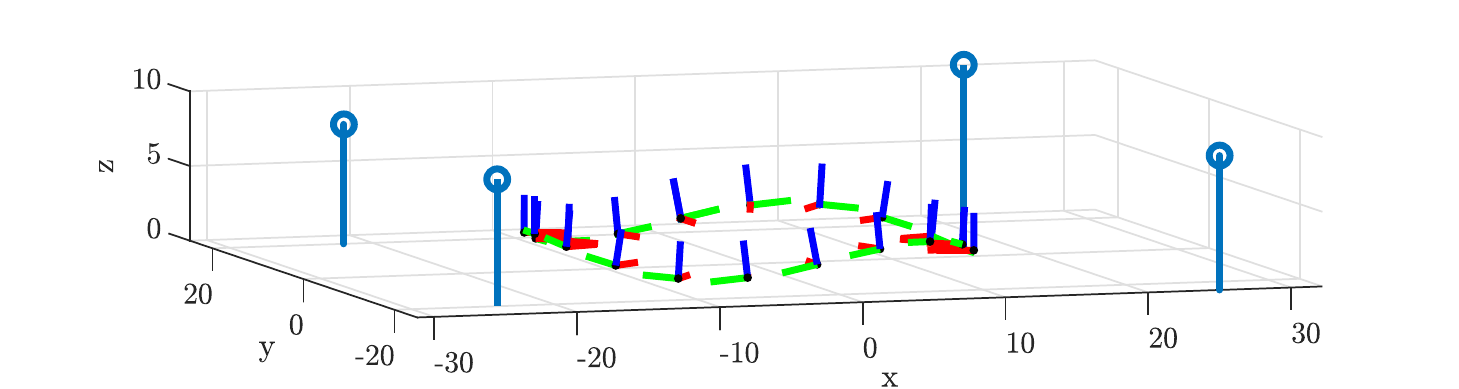}
	\caption{Simulation setup for the receiver. The three orthonormal vectors in three different colors (the red, green and blue vector represent the $x$-axis, $y$-axis and $z$-axis, respectively) at each sample on the path represent the frame of the receiver. The stems represent the BSs.}
	\label{fig:ch06env} \vspace{-5mm}
\end{figure}
To evaluate the performance of the proposed estimators, we consider a path with a circular pattern in the $xy$ plane and a sinusoidal pattern in the $z$ direction, as shown in Fig.~\ref{fig:ch06env}. The radius of the circle in the $xy$ plane is 15~m, and the amplitude of the sinusoid is 1~m. The circle is centered at $\left[0,0,1.5 \right]^{\operatorname{T}}$~m.
Starting at the coordinates $\left[0,15,1.5\right]^{\mathrm{T}}$~m, the path oscillates sinusoidally in the $z$ direction and completes the path in three periods.  

\begin{figure}
	\centering 
		\begin{subfigure}[t]{0.495\textwidth}
			\centering
			\includegraphics[width=\textwidth]{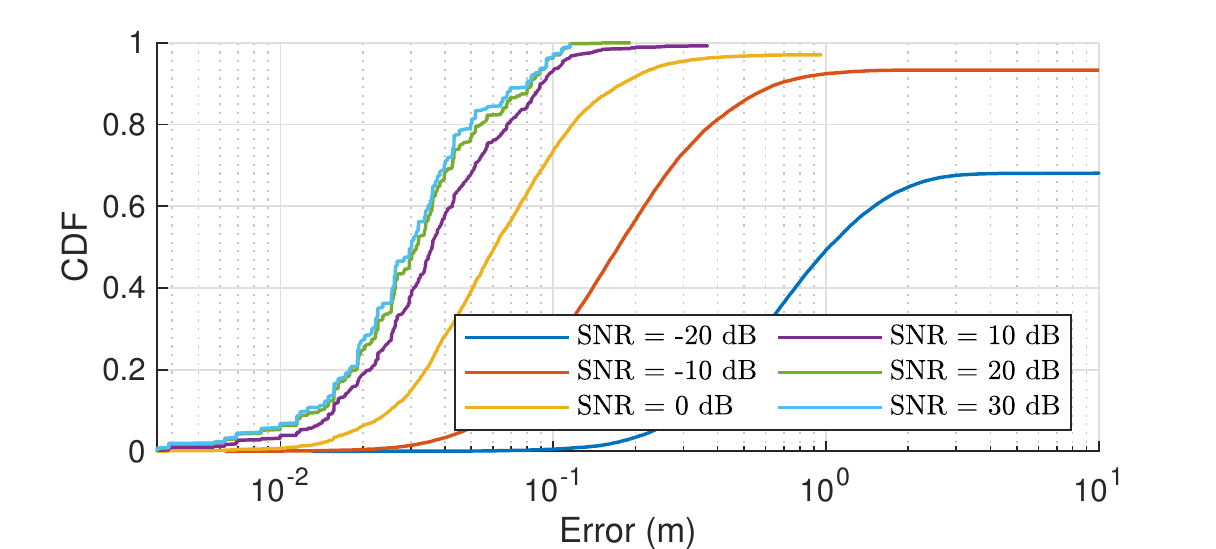}
			\caption{Position estimation error}%
			{{\small }}    
			\label{POS_CDFr}
		\end{subfigure}
		\begin{subfigure}[t]{0.495\textwidth}  
			\centering 
			\includegraphics[width=\textwidth]{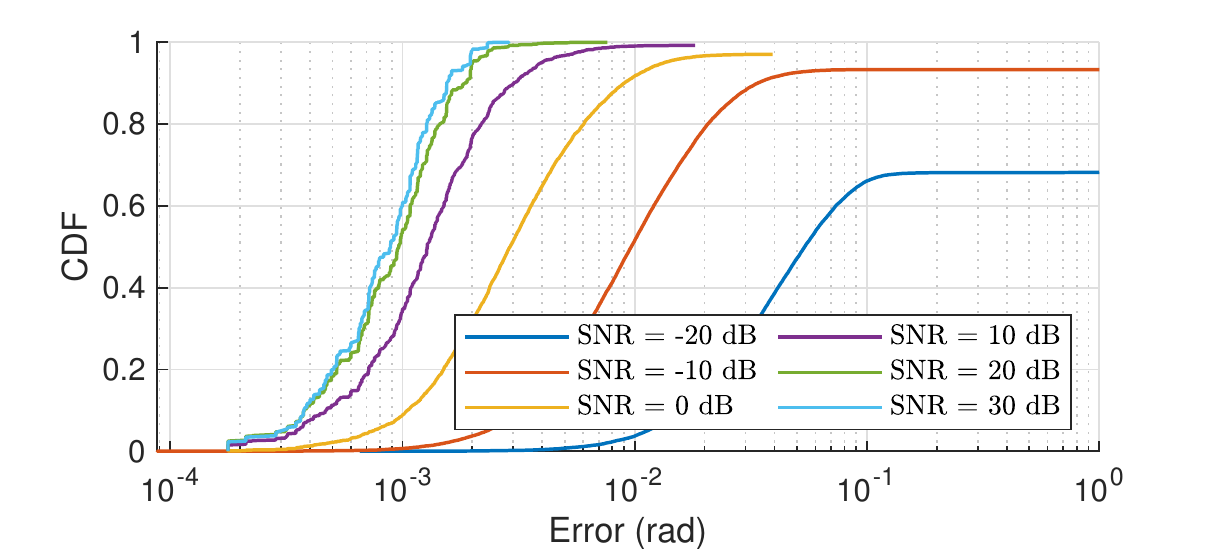}
			\caption[]{Orientation estimation error}%
			{{\small }}    
			\label{POS_CDFu}
		\end{subfigure}
	\caption[]
	{CDF of pose estimation errors for different \acp{SNR} for the \ac{AoA}-only pose estimation problem.}
	\label{fig:POS_CDF} \vspace{-5mm}
\end{figure}

\subsubsection{Results}
We evaluate the proposed estimator with respect to the SNR, and the pose estimate over the path  is evaluated. Fig.~\ref{fig:POS_CDF} shows the cumulative distribution (CDF) of the pose error. The figure reveals that our algorithm converges approximately for {95\%} of the sample positions when  $\text{SNR}=0$~dB, while this reduces to {70\%} when $\text{SNR}=-20$~dB. Hence, as expected, the coverage degrades if the receiver has a lower SNR. Furthermore, it can be seen that the performance of the estimator increases as the SNR increases, however, the performance gain decreases as the SNR increases.

\subsection{Problem 1 --  mmWave MIMO Snapshot SLAM: Establishing the Tightness of the Estimator}
In this section, we evaluate the asymptotic tightness of the proposed algorithm given by \eqref{Obj} by comparing it with the theoretical bound (See \cite{Shen19} for their derivation) under the assumption that the parameter vector estimate $\hat{\mathbf{z}}_{\ell}$ is given.
\subsubsection{Scenario}
\begin{figure}[!tb]
	\centering
	\includegraphics[width=0.495\textwidth]{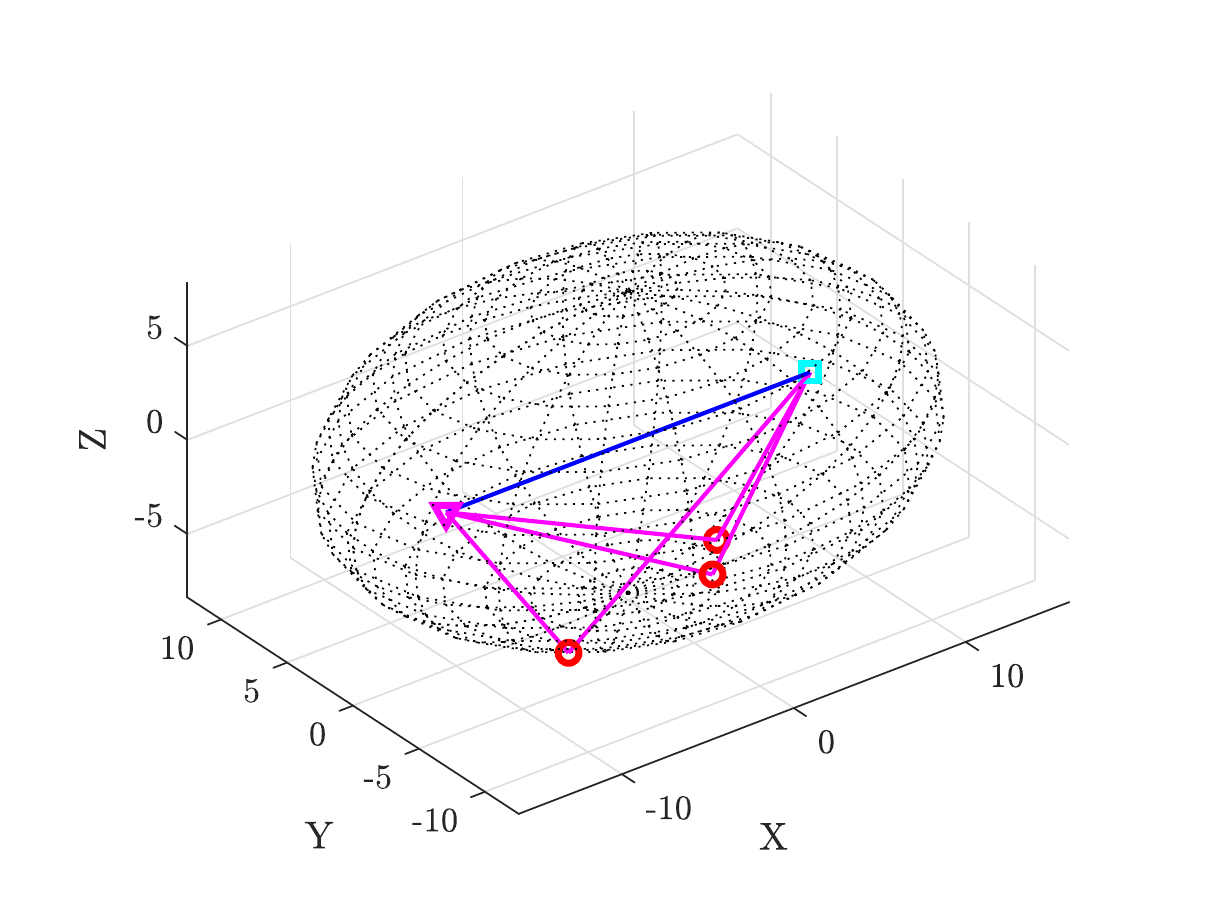}
	\caption{Simulation setup of 3D ellipsoid channel model, used to establish the tightness of the method from Section \ref{solutionProblem1} for the mmWave MIMO snapshot SLAM problem. The UE and BS (shown with a triangle and square) are the foci of the ellipsoid.}
	\label{fig:envellip}\vspace{-5mm}
\end{figure}

We first consider a channel model of 3D geometry-based stochastic model \cite{Hao21}, as shown in Fig.~\ref{fig:envellip}, where $N_s$ scatter points are randomly generated on the surface of an ellipsoid. The principal axes of the ellipsoid are set to $\left[16, 12, 8\right]^{\mathrm{T}}$~m. The UE and BS are placed at the foci and point towards each other. The field of view of the receiver is set to $4\pi/9$~rad. 
The observed parameter vector is assumed to be $\mathbf{y} = \bm{\mu} + \mathbf{w}_{\bm{\mu}}$ with $\mathbf{w}_{\bm{\mu}} \sim \mathcal{N}(\mathbf{0},\sigma^2_{\mu}\mathbf{\Sigma}\otimes\mathbf{I}_{N_s})$ and $\mathbf{\Sigma} = \mathrm{diag}(\left[1,1,1,1,10^{-15}\right]^{\mathrm{T}})$ with units 1 for the virtual points and $\mathrm{s}^2$ for the ToA, which is chosen to characterize the magnitude order relation between different components based on the MSE of estimates given by the Tensor-ESPRIT algorithm.

\begin{figure}
	\centering 
		\begin{subfigure}[t]{0.495\textwidth}
			\centering
			\includegraphics[width=\textwidth]{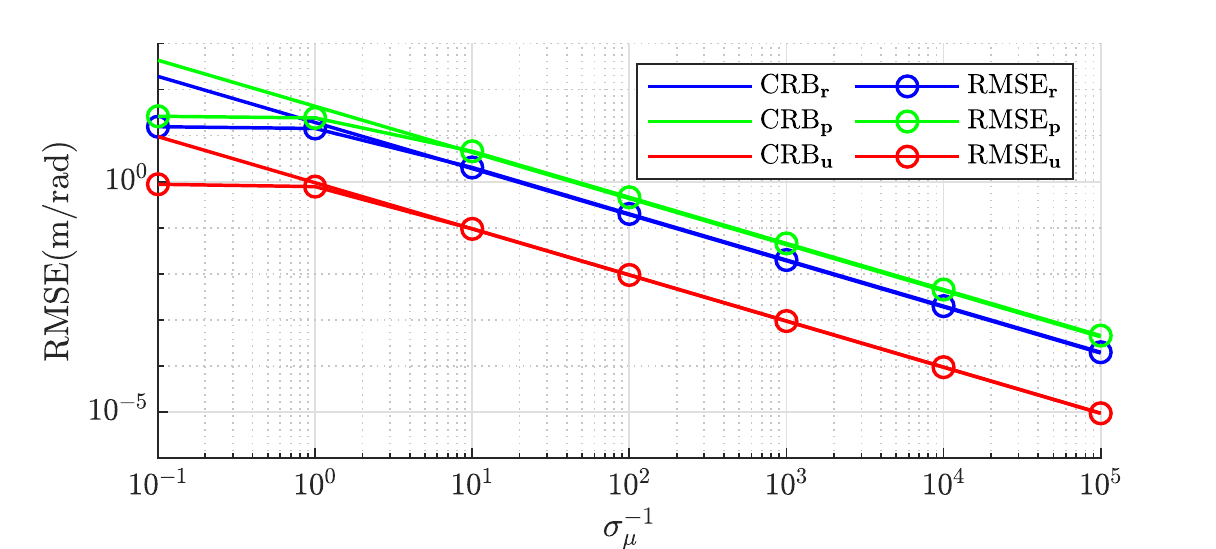}
			\caption[]%
			{ RMSE as a function of SNR $\sigma_{\mu}^{-1}$.}    
			\label{Tightness_SNR}
		\end{subfigure}
		\begin{subfigure}[t]{0.495\textwidth}  
			\centering 
			\includegraphics[width=\textwidth]{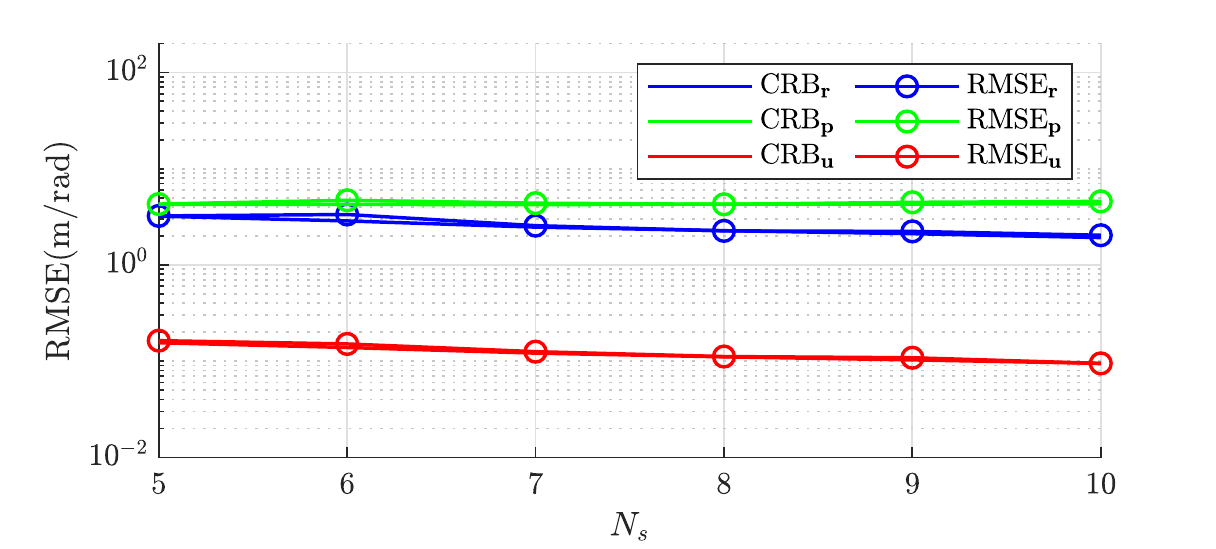}
			\caption[]%
			{RMSE as a function of number of scatter points $N_s$.}    
			\label{Tightness_NUM}
		\end{subfigure}
	\caption[]
	{RMSE for position, orientation, and the position of scatters as a function of (a) $\sigma_{\mu}^{-1}$ and (b) $N_s$.}
	\label{fig_Tightness} \vspace{-5mm}
\end{figure}

\subsubsection{Results}
First, in Fig.~\ref{Tightness_SNR}, we show the RMSE of the closed-form (denoted by CF) and the iterative algorithm (denoted by LS) solutions for the position,  orientation, and scatter positions as a function of the $\sigma_{\mu}^{-1}$ for $N_s=10$ scatter points. For the lease squares method, the weight of the $\ell^{\text{th}}$ path is set to $\mathbf{w}_{\ell} = |\alpha_{\ell}|^2[1,1,1,1,10^{15}]^{\mathrm{T}}$. Note that since the association between the estimated and true scatters cannot be recovered, the association for the scatters is chosen such that the estimation error for each scatter is minimized when calculating the RMSE. The results show that the proposed estimator is indeed asymptotically tight for large SNR. There, the RMSE for all types of errors reduces in inverse proportion to the SNR, i.e., the proposed algorithm is able to estimate the UE pose and the position of scatters with high accuracy from the observed parameter vector. Next, we show in Fig.~\ref{Tightness_NUM} the RMSE for the position, orientation, and scatter positions as a function of the number $N_s$ of scatters, for $\sigma^{-1}_{\mu}=10$. Also here, we see an tightness between the RMSEs and their respective lower bounds, however, the performance remains almost constant with respect to $N_s$. This is expected because although the number of observations of the estimator increases with the number of scatter points, the number of parameters to be estimated also increases.
\subsection{Problem 1 --  mmWave MIMO Snapshot SLAM: In-depth Performance Evaluation}
In this section, we evaluate the estimator in the simulation setting given by \cite{Wen21}. 
\subsubsection{Scenario}
\begin{figure}[!tb]
	\centering
	\includegraphics[width=0.495\textwidth]{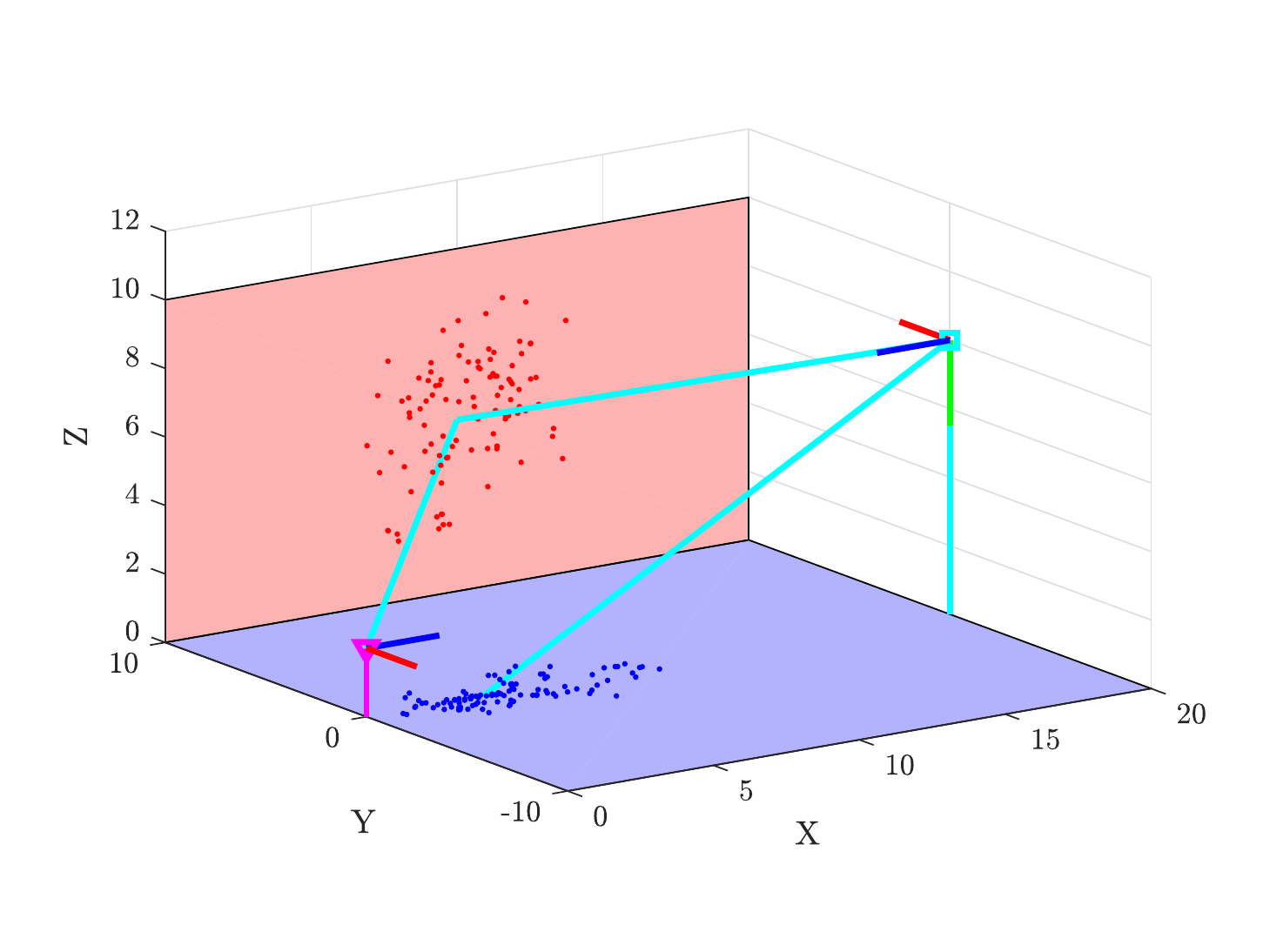}
	\vspace{-12mm}
	\caption{Simulation setup for channel estimation and SLAM with 2 clusters.}
	\label{fig:envslam}\vspace{-5mm}
\end{figure}

We consider a scene consisting of two surfaces, representing the building and ground surfaces respectively, as shown in Fig.~\ref{fig:envslam}. The building facade's center is at $\left[10,10,5\right]^{\mathrm{T}}$~m with facade length of 20 m, facade height of 10 m, and direction $\left[0,1,0\right]^{\mathrm{T}}$. The ground surface is at $\left[10,0,0\right]^{\mathrm{T}}$~m with direction $\left[0,0,1\right]^{\mathrm{T}}$, surface dimension $20\times 20$~m$^2$. The positions of BS and UE are set to $\mathbf{r}_{\mathrm{BS}}=\left[20,0,8\right]^{\mathrm{T}}$~m and $\mathbf{r}_{\mathrm{UE}}=\left[0,0,2\right]^{\mathrm{T}}$~m, respectively, while their orientations are 
\begin{align}
 \mathbf{R}_{\mathrm{BS}} = \begin{bmatrix}
0,&0,&-1\\1,&0,&0\\0,&-1,&0
\end{bmatrix}, \quad \mathbf{R}_{\mathrm{UE}}=\begin{bmatrix}
0,&0,&1\\-1,&0,&0\\0,&-1,&0
\end{bmatrix},   
\end{align}
pointing at each other.
The BS and UE are equipped with a uniform rectangular array with $10\times 10$ half-wavelength spaced elements and a carrier frequency of 28 GHz.
For the $k^{\text{th}}$ facade, $L_{k}=100$ scatters are generated ($L = 200$ in total), whose positions $\mathbf{p}_{\ell}$ are randomly generated according to the model from \cite{Wen215,Wen21,Yu20,kulmer2019high}, described in Appendix \ref{app:Generative-Model}.  This model is parameterized by $\beta \in \mathbb{N}$, which describes the directivity of the scattering (i.e., larger $\beta$ means more directive scattering). 
Following \cite{Wen21,Wen215}, we assume that the received pilot signal at $n^{\text{th}}$ subcarrier is 
\begin{equation}
    \bm{\mathsf{Y}}[n] = \bm{\mathsf{H}}[n]\bm{\mathsf{X}}[n] + \bm{\mathsf{W}}[n],
\end{equation}
where $\bm{\mathsf{W}}[n]$ models the additive Gaussian noise, and $\bm{\mathsf{X}}[n]$ is chosen such that $(\bm{\mathsf{X}}[n])(\bm{\mathsf{X}}[n])^{\mathrm{H}} = \mathbf{I}$. Then, the channel estimate is given by $\hat{\bm{\mathsf{H}}}[n] = \bm{\mathsf{Y}}[n](\bm{\mathsf{X}}[n])^{\mathrm{H}}$ \cite{Wen21}, and after stacking and rearranging the channel estimates of all $N_f$ subcarriers, we have a tensor representation of the channel estimates, given by
\begin{equation}\label{tensorH}
    \hat{\bm{\mathcal{H}}} = \bm{\mathcal{H}} + \bm{\mathcal{W}} \in \mathbb{C}^{N_{r,x}\times N_{r,y}\times N_{t,x}\times N_{t,y}\times N_f}.
\end{equation}
In \eqref{tensorH}, the tensor $\bm{\mathcal{H}}$ represents the true value of the channel, and the tensor $\bm{\mathcal{W}}$ represents the estimation error contained in $\hat{\bm{\mathcal{H}}}$, which is Gaussian. The SNR is defined as SNR $=\|\bm{\mathcal{H}}\|^2_F/\|\bm{\Sigma}_{\bm{\mathcal{H}}}\|_F$ with $\bm{\Sigma}_{\bm{\mathcal{H}}}$ the covariance of $\bm{\mathcal{W}}$. The MATLAB package
Tensorlab \cite{vervliet2016tensorlab} is used to perform tensor-ESPRIT algorithm \cite{Wen21,Wen215} to estimate the parameter vector set $\mathcal{Z}$ from $\hat{\bm{\mathcal{H}}}$. The number of paths is estimated by the minimum description length (MDL) \cite{Yokota17}, this estimated number $\hat{L}$ is usually much less than $L$, representing the number of the effective scattering paths. Subsequently, the proposed iterative method is applied to the estimated $\mathcal{Z}$ to implement SLAM and initialized by the closed-form solution.

\begin{figure*}
	\centering 
		\begin{subfigure}[t]{0.495\textwidth}
			\centering
			\includegraphics[width=\textwidth]{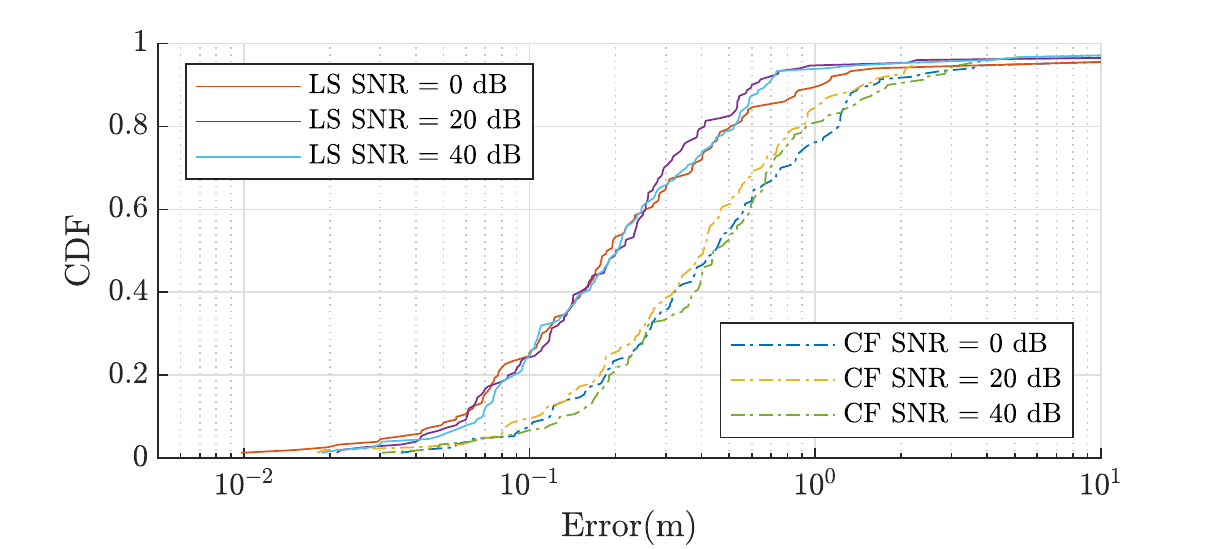}
			\caption[]%
			{{\small }}    
			\label{SLAM_aSYN_LOS_SNR_CDFr}
		\end{subfigure}
		\hfill
		\begin{subfigure}[t]{0.495\textwidth}  
			\centering 
			\includegraphics[width=\textwidth]{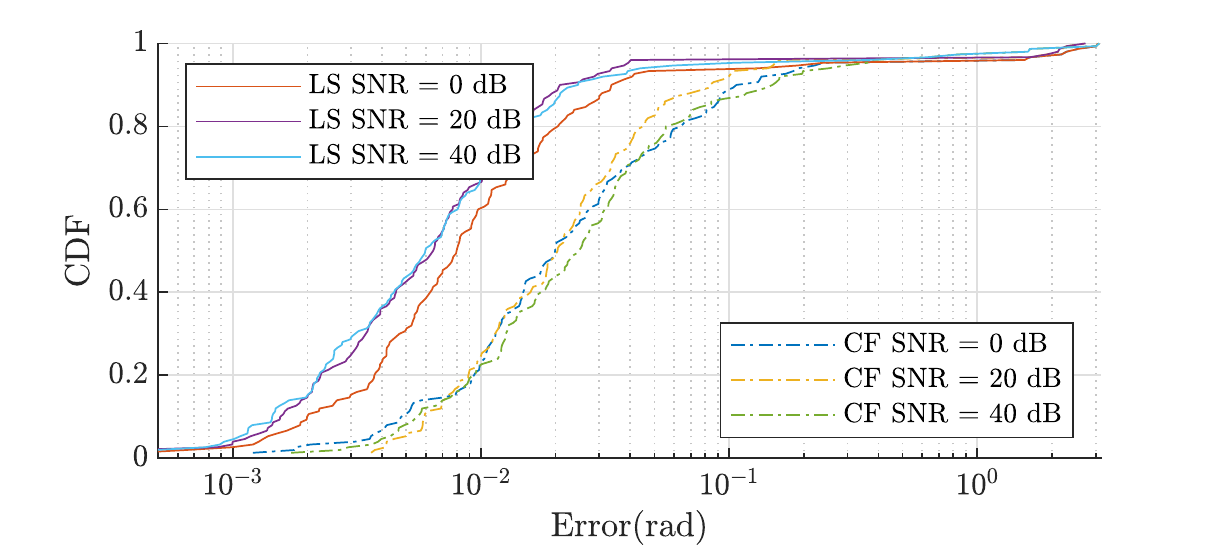}
			\caption[]%
			{{\small }}    
			\label{SLAM_aSYN_LOS_SNR_CDFu}
		\end{subfigure}
		\vfill
		\begin{subfigure}[t]{0.495\textwidth}
			\centering
			\includegraphics[width=\textwidth]{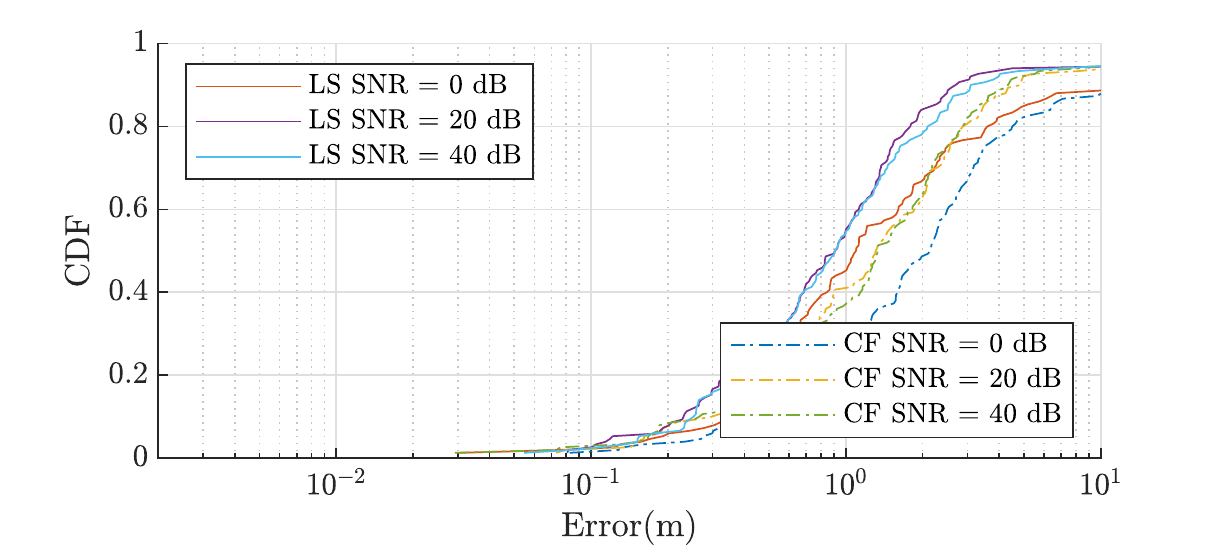}
			\caption[]%
			{{\small }}    
			\label{SLAM_aSYN_NLOS_SNR_CDFr}
		\end{subfigure}
		\hfill
		\begin{subfigure}[t]{0.495\textwidth}  
			\centering 
			\includegraphics[width=\textwidth]{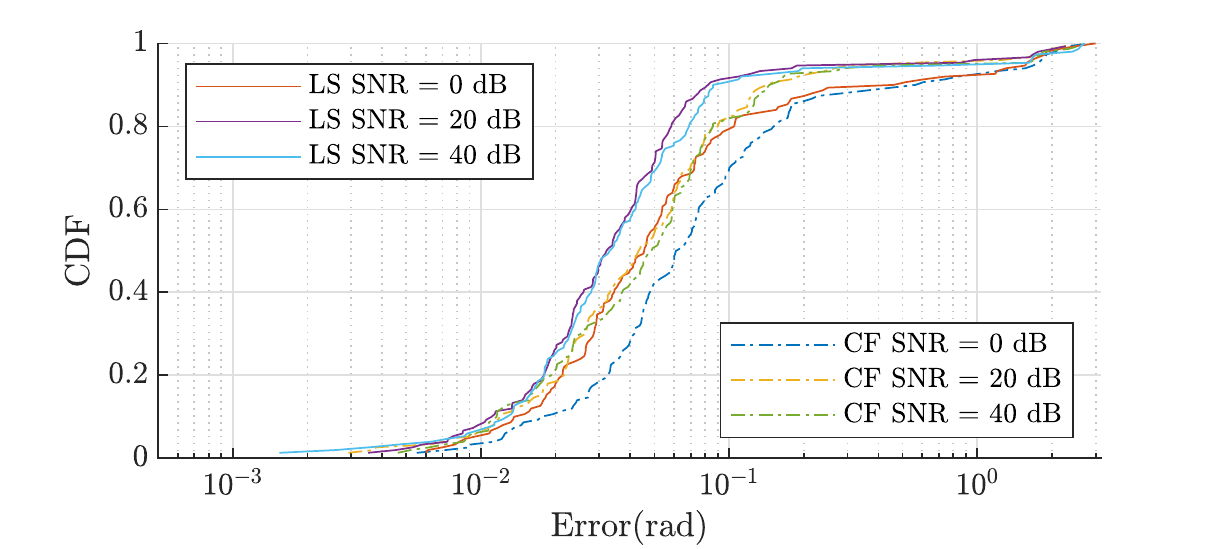}
			\caption[]%
			{{\small }}    
			\label{SLAM_aSYN_NLOS_SNR_CDFu}
		\end{subfigure}
	\caption[]
	{CDF of pose errors for different SNRs for the closed form (CF) and least squares (LS) solution. The position error is given in (a) and (c), while the orientation error is given in (b) and (d). Top and bottom rows contain the performance in the \ac{LoS} and \ac{NLoS} cases, respectively.}
	\label{fig:SLAM_aSYN_SNR_CDF} \vspace{-5mm}
\end{figure*}

\begin{figure*}
	\centering 
		\begin{subfigure}[t]{0.495\textwidth}
			\centering
			\includegraphics[width=\textwidth]{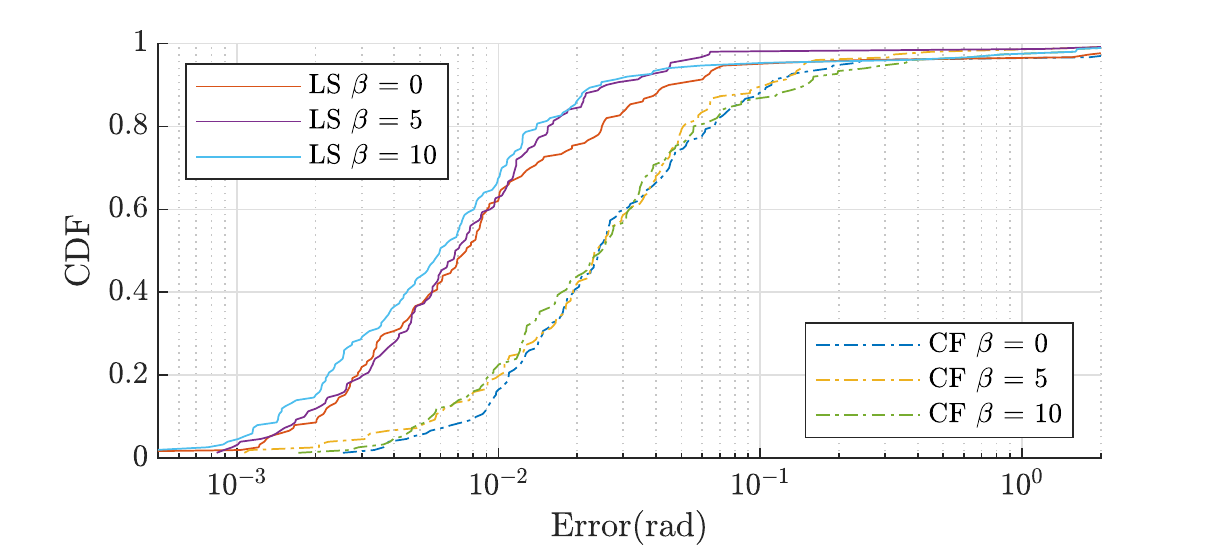}
			\caption[]%
			{{\small }}    
			\label{SLAM_aSYN_LOS_Alf_CDFr}
		\end{subfigure}
		\hfill
		\begin{subfigure}[t]{0.495\textwidth}  
			\centering 
			\includegraphics[width=\textwidth]{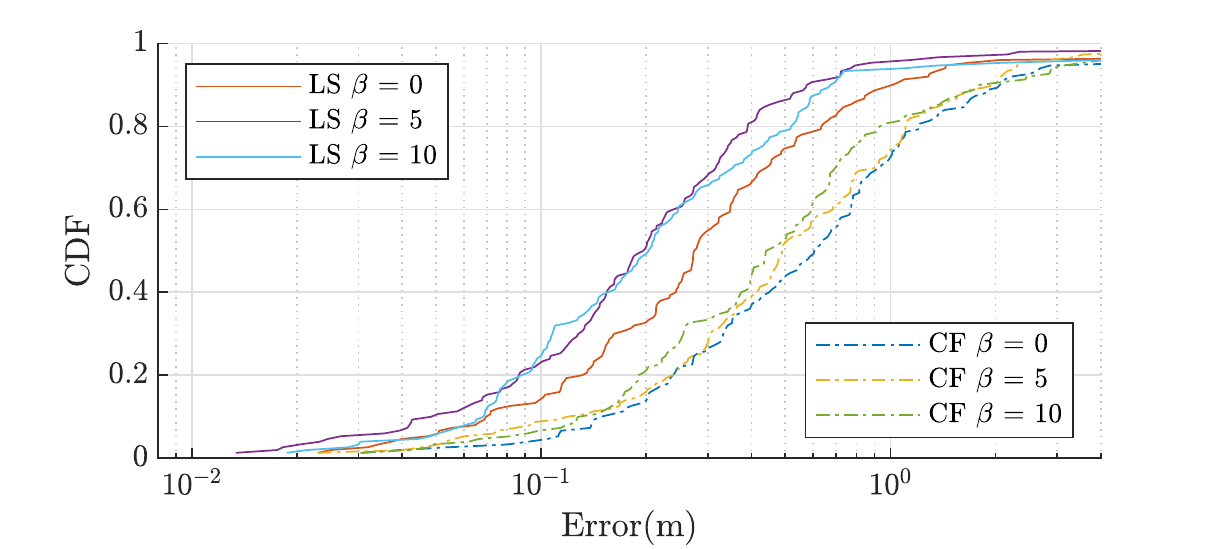}
			\caption[]%
			{{\small }}    
			\label{SLAM_aSYN_LOS_Alf_CDFu}
		\end{subfigure}
		\vfill
		\begin{subfigure}[t]{0.495\textwidth}
			\centering
			\includegraphics[width=\textwidth]{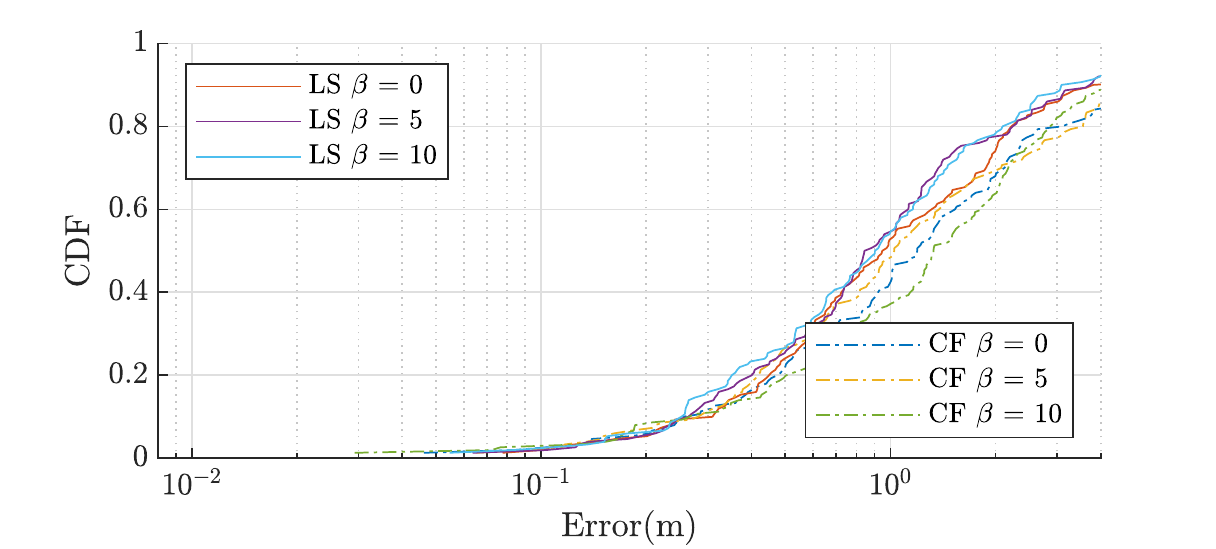}
			\caption[]%
			{{\small }}    
			\label{SLAM_aSYN_NLOS_Alf_CDFr}
		\end{subfigure}
		\hfill
		\begin{subfigure}[t]{0.495\textwidth}  
			\centering 
			\includegraphics[width=\textwidth]{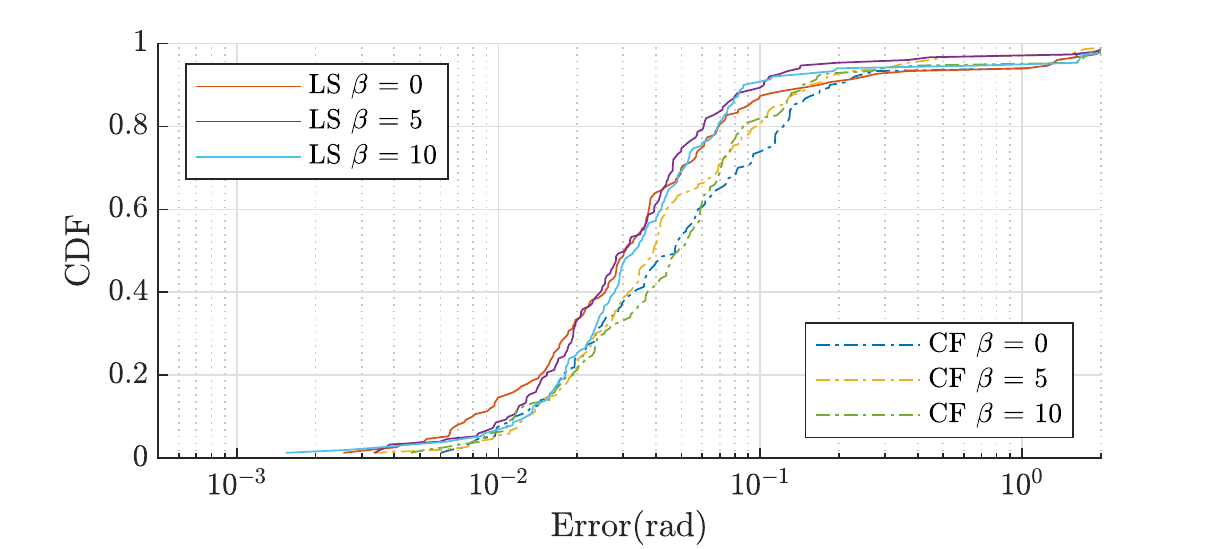}
			\caption[]%
			{{\small }}    
			\label{SLAM_aSYN_NLOS_Alf_CDFu}
		\end{subfigure}
	\caption[]
	{CDF of pose errors for different values of the directivity $\beta$ for the closed form (CF) and least squares (LS) solution. The position error is given in (a) and (c), while the orientation error is given in (b) and (d). Top and bottom rows contain the performance in the \ac{LoS} and \ac{NLoS} cases, respectively.}
	\label{fig:SLAM_aSYN_Alf_CDF} \vspace{-5mm}
\end{figure*}

\subsubsection{Results}
First, in Fig.~\ref{fig:SLAM_aSYN_SNR_CDF}, we show the CDF of the estimation errors, for different SNRs for $\beta=10$. The performance of both closed-form estimation and LS estimation is included and denoted as CF and LS, respectively, for comparison. Under the condition of asynchronization, top and bottom rows of Fig.~\ref{fig:SLAM_aSYN_SNR_CDF} contain the performance in the \ac{LoS} and \ac{NLoS} cases, respectively.
We observe that the LS algorithm outperforms the CF method, in both the \ac{LoS} and \ac{NLoS} cases. 
The figure further reveals that 
the LS algorithm achieves an accuracy of 1~m and 0.03~rad or better in approximately 90\% of the cases when the \ac{LoS} channel is present, while this accuracy becomes 3~m and 0.1~rad in approximately 90\% of the cases when the \ac{LoS} is absent. 
It can also be seen that the performance does not improve significantly with increasing SNR, which means that SNR is not the dominant factor in determining performance. This is due to the fact that both the LS and CF algorithms utilize the estimated parameter set $\tilde{\mathcal{Z}}$, which is provided in the form of effective scatter points and has limited resolution, so the performance is mainly limited by the accuracy of $\tilde{\mathcal{Z}}$. 
Next, we show in Fig.~\ref{fig:SLAM_aSYN_Alf_CDF} the CDF of the estimation errors for different values of the directivity $\beta$, for SNR $=40$~dB. Also here, we see an outperforming of the LS method against the CF method. 
The figure further reveals that 
the LS algorithm achieves an accuracy of 1~m and 0.03~rad or better in approximately 90\% of the cases when the \ac{LoS} channel is present, while this accuracy becomes 3~m and 0.1~rad in approximately 90\% of the cases when the \ac{LoS} is absent. 
The results shown in Fig.~\ref{fig:SLAM_aSYN_Alf_CDF} reflect that $\beta$ has a smaller impact on performance when the \ac{LoS} channel is present.

\section{Conclusion}
In this paper, we investigate the estimation of the full 6D user pose (joint 3D position and 3D orientation) using antenna arrays by providing a projective geometric view of \acp{AoD} and \acp{AoA} in the context of 5G and beyond 5G positioning. To this end, the directional angular information is first modeled in terms of the receiver's pose using the perspective projection model from computer vision. Then, two pose estimation  problems, namely 6D pose estimation using \acp{AoA} from multiple base stations and 6D SLAM based on single-BS mmWave communication, are investigated with the perspective projection model. Particularly, we show that the 6D SLAM problem, when modeled with the perspective projection model, can be further modeled with the epipolar model. For each problem, we propose two estimation algorithms, a closed-form one and an iterative one based on the principle of \ac{LS}. The simulation results confirm the effectiveness of the proposed algorithms and demonstrate that reliable 6D localization of a user is achievable even in the absence  of the \ac{LoS} path, while the performance is significantly improved when the \ac{LoS} path is present.

\appendices

\section{Brief introduction to the five point algorithm for estimating the essential matrix}\label{sect:append_5pt}
Base on \cref{epiE}, for a set of five correspondences $\{(\bar{\mathbf{v}}_i,\bar{\bm{\nu}}_i)\}$, we have
\begin{equation}\label{fpa}
\underbrace{\begin{bmatrix}
    \bar{\mathbf{v}}_1^{\mathrm{T}}\otimes\bar{\bm{\nu}}_1^{\mathrm{T}} \\
    \vdots \\
    \bar{\mathbf{v}}_5^{\mathrm{T}}\otimes\bar{\bm{\nu}}_5^{\mathrm{T}}
\end{bmatrix}}_{\mathbf{A}\in\mathbf{R}^{5\times 9}}
\mathrm{vec}(\mathbf{E}) = \mathbf{0}
\end{equation}
where $\mathrm{vec}(\mathbf{E}) = \left[\bm{\mathfrak{e}}_1^{\mathrm{T}},\bm{\mathfrak{e}}_2^{\mathrm{T}},\bm{\mathfrak{e}}_3^{\mathrm{T}}\right]^{\mathrm{T}}$. \cref{fpa} implies that $\mathrm{vec}(\mathbf{E})$ is the right null space of $\mathbf{A}$, whose base vectors $\mathrm{vec}(\mathbf{E}_i)$, $i\in\{1,\dots 4\}$ can be obtained with SVD. Since $\mathbf{E}$ is defined up to a scale, we have $\mathbf{E} = x\mathbf{E}_1+y\mathbf{E}_2+z\mathbf{E}_3+\mathbf{E}_4$, for certain coefficients $\mathcal{C} = (x,y,z)$. Further, the following constraints,   
\begin{equation}
    \det(\mathbf{E}) = 0
\end{equation}
and
\begin{equation}
    \mathbf{E}\mathbf{E}^{\mathrm{T}}\mathbf{E} -\frac{1}{2}\mathrm{tr}(\mathbf{E}\mathbf{E}^{\mathrm{T}})\mathbf{E} = \mathbf{0}_{3},
\end{equation}
are used to build 10 third-order polynomial equations in $\mathcal{C}$. In general, $K \leq 10$ sets of real solutions to $\mathcal{C}$ are obtained from the polynomial equations. Since the five point algorithm is used together with the RANSAC strategy, in order to keep only the optimal solution, these $K$ solutions are assessed with the remaining correspondences other than the chosen five correspondences.

\section{Estimation with the knowledge of the \ac{LoS} path}
\label{sect:append_los}
Prior knowledge of the existence of the \ac{LoS} path can improve the performance of SLAM, since the \ac{AoD} and \ac{AoA} associated with the \ac{LoS} channel are specifically represented by epipoles in epipolar geometry.
For the sake of simplification, we ignore the subscript
of $\mathbf{R}_{\mathrm{UE}}$ in this section.
As pointed out in \cite[Sect. 2.2]{faugeras1990motion}, the epipoles $\bm{\nu}_0$ and $\mathbf{v}_0$ have the following two properties
\begin{align}
    \bar{\bm{\nu}}_0 &  = \mathbf{r}_{\mathrm{UE}} \label{lepi1}\\
    \bar{\mathbf{v}}_0 & = - \mathbf{R}^{\mathrm{T}} {\mathbf{r}}_{\mathrm{UE}},\label{lepi2}
\end{align}
where these two equalities are defined up to  a scale. 
These two properties can be exploited to recover the pose partially. On the one hand, the direction of $\mathbf{r}_{\mathrm{UE}}$ can be recovered according to \cref{lepi1}. On the other hand, \cref{lepi2} implies that the rotated vector $\mathbf{R}\bar{\mathbf{v}}_0$ specifies the direction of  $-\bar{\bm{\nu}}_0$, with which we can recover $\mathbf{R}$ partially. To illustrate this, substituting \cref{lepi1}
into \cref{lepi2} and left multiplying $\mathbf{R}$, we have, up to  a scale, 
\begin{equation}\label{lepi3}
    -\bar{\bm{\nu}}_0=\mathbf{R}\bar{\mathbf{v}}_0.
\end{equation}
Then following the swing-twist parameterization \cite[Sect. 5]{grassia1998practical}, the rotation $\mathbf{R}$ is decomposed into $\mathbf{R} = \mathbf{R}_{\perp}\mathbf{R}_{\parallel}$, where $\mathbf{R}_{\parallel} = \exp\left(\theta(\mathbf{n}_{\parallel})_{\times}\right)$ is a rotation matrix over an arbitrary unit vector  $\mathbf{n}_{\parallel}$ and an arbitrary angle $\theta$, and $\mathbf{R}_{\perp} = \exp\left((\mathbf{u}_{\perp})_{\times}\right)$ is the rotation matrix of a rotation vector $\mathbf{u}_{\perp}$ residing in the plane perpendicular to $\mathbf{n}_{\parallel}$, i.e., $\mathbf{u}_{\perp}^{\mathrm{T}}\mathbf{n}_{\parallel} = 0$. Note that for arbitrary $\theta$ we have $\mathbf{n}_{\parallel} = \mathbf{R}_{\parallel}\mathbf{n}_{\parallel}$. Then if we choose $\mathbf{n}_{\parallel} = \tfrac{\bar{\mathbf{v}}_0}{\|\bar{\mathbf{v}}_0\|}$,  for arbitrary $\theta$, \cref{lepi3} can be rewritten as 
\begin{eqnarray}
    -\bar{\bm{\nu}}_0 &=& \mathbf{R}_{\perp}\mathbf{R}_{\parallel} \bar{\mathbf{v}}_0 \\
    &=& \mathbf{R}_{\perp}\bar{\mathbf{v}}_0. \label{vRv}
\end{eqnarray}
Since $\mathbf{R}_{\perp}$ brings $\bar{\mathbf{v}}_0$ to the direction of $-\bar{\bm{\nu}}_0$, the axis of rotation $\mathbf{R}_{\perp}$ is therefore $\frac{\bar{\bm{\nu}}_0\times\bar{\mathbf{v}}_0 }{\|\bar{\bm{\nu}}_0\times\bar{\mathbf{v}}_0\|}$ and the rotation angle is the angle between $\bar{\mathbf{v}}_0$ and $-\bar{\bm{\nu}}_0$, then we have \cite[Eq. 3]{kallmann2008analytical}
\begin{align}
    \mathbf{u}_{\perp} = \frac{\bar{\bm{\nu}}_0 \times\bar{\mathbf{v}}_0 }{\|\bar{\bm{\nu}}_0\times\bar{\mathbf{v}}_0\|}\arccos{\tfrac{-\bar{\mathbf{v}}_0^{\mathrm{T}}\bar{\bm{\nu}}_0}{\|\bar{\mathbf{v}}_0\|\|\bar{\bm{\nu}}_0\|}},
\end{align}
which has two degrees of freedom \cite[Sect. 5]{grassia1998practical}. %
It can be seen that based merely on the \ac{AoA} and \ac{AoD} of the \ac{LoS} path, two degrees of freedom specified by $\mathbf{u}_{\perp}$ can be recovered, but not the remaining one, $\theta$ (i.e.,  $\mathbf{R} = \exp\left((\mathbf{u}_{\perp})_{\times}\right)\exp\left(\theta(\mathbf{n}_{\parallel})_{\times}\right)$ where only $\theta$ remains unknown). To this end, we use the information 
provided by the scattering paths. Based on the facts that $ \bar{\bm{\nu}}^{\mathrm{T}}\mathbf{E}\bar{\mathbf{v}} = 0$ from \eqref{epiE}, $\mathbf{E} = (\mathbf{r}_{\mathrm{UE}})_{\times}\mathbf{R}$ from \eqref{ErR} and the relation \eqref{lepi1}, an additional pair of correspondences $(\bm{\nu}_l,\mathbf{v}_l)$ associated with the $l^{\text{th}}$ scatter point leads to
\begin{align}
      &\bar{\bm{\nu}}_l^{\mathrm{T}}(\bar{\bm{\nu}}_{0})_{\times}\mathbf{R}\bar{\mathbf{v}}_l  = \bar{\bm{\nu}}_l^{\mathrm{T}}(\bar{\bm{\nu}}_{0})_{\times}\mathbf{R}_{\perp}\mathbf{R}_{\parallel}\bar{\mathbf{v}}_l = \bm{\mathfrak{v}}_l^{\mathrm{T}}\exp(\theta(\mathbf{n}_{\parallel})_{\times})\bar{\mathbf{v}}_l \nonumber  \\
      =& \bm{\mathfrak{v}}_l^{\mathrm{T}}\bar{\mathbf{v}}_l + \bm{\mathfrak{v}}_l^{\mathrm{T}}(\mathbf{n}_{\parallel})_{\times}\bar{\mathbf{v}}_l\sin\theta+\bm{\mathfrak{v}}_l^{\mathrm{T}}(\mathbf{n}_{\parallel})^2_{\times}(1-\cos\theta)\bar{\mathbf{v}}_l \label{Rod}\\
      =& \bm{\mathfrak{v}}_l^{\mathrm{T}}(\mathbf{n}_{\parallel})_{\times}\bar{\mathbf{v}}_l\sin\theta + \bm{\mathfrak{v}}_l^{\mathrm{T}}\bar{\mathbf{v}}_l\cos\theta \label{Cros}\ \\
      = & \bm{\mathfrak{n}}_l^{\mathrm{T}}
    \begin{bmatrix}
    \cos(\theta) \\
    \sin(\theta) \\
    1
    \end{bmatrix} 
    =0. 
\end{align}
where $\bm{\mathfrak{v}}_l^{\mathrm{T}} = \bar{\bm{\nu}}_l^{\mathrm{T}}(\bar{\bm{\nu}}_{0})_{\times}\mathbf{R}_{\perp}$, and
\begin{equation}
    \bm{\mathfrak{n}}_l=
    \begin{bmatrix}
    \bm{\mathfrak{v}}_l^{\mathrm{T}}\bar{\mathbf{v}}_l \\
    \bm{\mathfrak{v}}_l^{\mathrm{T}}(\mathbf{n}_{\parallel}\times\mathbf{v}_l) \\
    0
    \end{bmatrix}.
\end{equation}
Eq. (\ref{Rod}) holds due to the Rodrigues' rotation formula, given by,
\begin{equation}
    \exp(\theta\mathbf{n}_{\times}) = \mathbf{I} + \mathbf{n}_{\times}\sin\theta+\mathbf{n}^2_{\times}(1-\cos\theta),
\end{equation}
while \eqref{Cros} holds due to the cross product properties of $\mathbf{a}\times\mathbf{a} = \mathbf{0}$ and $\mathbf{a}\times(\mathbf{b}\times\mathbf{c}) = (\mathbf{a}^{\mathrm{T}}\mathbf{c})\mathbf{b} - (\mathbf{a}^{\mathrm{T}}\mathbf{b})\mathbf{c}$ and to the equality $\bm{\mathfrak{v}}_l^{\mathrm{T}}\mathbf{n}_{\parallel} = 0$ resulted from (\ref{vRv}). It can be seen that $\theta$ corresponds to the intersections of the unit circle centered at origin with the line specified by the normal vector $\bm{\mathfrak{n}}_l$ and the origin. There are two intersections, thus two values $\theta_i$, $i\in \{1,2\}$, which lead to two estimates of $\mathbf{R}$. 
By first noticing that the rotated vector $\mathbf{R}\bar{\mathbf{v}}_l$ resides in the epipolar plane and that
\begin{eqnarray}
    (\mathbf{R}|_{\theta_1}\bar{\mathbf{v}}_l)\times(\mathbf{R}|_{\theta_1}\mathbf{n}_{\parallel}) &=& (\mathbf{R}|_{\theta_2}\mathbf{n}_{\parallel})\times(\mathbf{R}|_{\theta_2}\bar{\mathbf{v}}_l), \\
    \mathbf{R}|_{\theta_1}\mathbf{n}_{\parallel} &=& \mathbf{R}|_{\theta_2}\mathbf{n}_{\parallel},
\end{eqnarray}
where $\mathbf{R}|_{\theta} \doteq \mathbf{R}_{\perp}\exp(\theta(\mathbf{n}_{\parallel})_{\times})$, then the two estimates can be illustrated by the vectors  $\mathbf{R}|_{\theta_1}\bar{\mathbf{v}}_l$ and $\mathbf{R}|_{\theta_2}\bar{\mathbf{v}}_l$ forming reflection with respect to $\mathbf{R}\mathbf{n}_{\parallel}$ within the epipolar plane. As a result, we can choose the appropriate value for $\theta$ so that $\bar{\bm{\nu}}_0\times\bar{\bm{\nu}}_l$ and $\bar{\bm{\nu}}_0\times(\mathbf{R}\bar{\mathbf{v}}_l)$ have the identical sign.
Thus, one additional scattering point suffices to uniquely recover the rotation matrix. 

The sequential application of \cref{lepi1} and (\ref{Rod}) can be used to estimate the pose up to a scale factor. However, in this approach, the direction estimation of $\mathbf{r}_{\mathrm{UE}}$ is based exclusively on the observation of the \ac{LoS} channel, while the observation of scattering paths does not contribute to this estimation. To deal with this problem and to make the estimation compatible with existing algorithms, we can convert the constraints \cref{lepi1} and \cref{lepi2} into a set of extra correspondences
$\mathcal{S} = \{(\mathbf{e}_1,\bar{\mathbf{v}}_0), (\mathbf{e}_2,\bar{\mathbf{v}}_0),  (\mathbf{e}_3,\bar{\mathbf{v}}_0),(\bar{\bm{\nu}}_0,\mathbf{e}_1),(\bar{\bm{\nu}}_0,\mathbf{e}_2),(\bar{\bm{\nu}}_0,\mathbf{e}_3)\}$, which can then be fed into existing algorithms along with the virtual points of scattering paths.

\section{Gradients in the Gauss-Newton Method}\label{app:gradients}
The gradient of $\tau_{\ell}$ with respect to $\bm{\Theta}$ is given by
\begin{equation}
\nabla_{\bm{\Theta}}\tau_{\ell} = \frac{1}{c}\Big(\frac{\mathbf{p}_{\ell}^{\mathrm{T}}}{\|\mathbf{p}_{\ell}\|}\nabla_{\bm{\Theta}}\mathbf{p}_{\ell} + \frac{\mathbf{p}_{\ell,u}^{\mathrm{T}}}{\|\mathbf{p}_{\ell,u}\|} \nabla_{\bm{\Theta}}\mathbf{p}_{\ell,u}\Big),
\end{equation}
where $\nabla_{\bm{\Theta}}\mathbf{p}_{\ell} = \left[\nabla_{\mathbf{T}_{\mathrm{UE}}}^{\mathrm{T}}\mathbf{p}_{\ell}, \nabla_{\mathbf{p}_{1}}^{\mathrm{T}}\mathbf{p}_{\ell},\dots,\nabla_{\mathbf{p}_L}^{\mathrm{T}}\mathbf{p}_{\ell}\right]$, 
$\nabla_{\mathbf{T}_{\mathrm{UE}}}\mathbf{p}_{\ell} = \mathbf{0}_{3\times 6}$,
\begin{eqnarray}
\nabla_{\mathbf{p}_i}\mathbf{p}_{\ell} = \left\{
\begin{array}{cll}
\mathbf{I}_3 & \quad & \text{if } \ell = i \\
\mathbf{0}_{3\times3}                                           & \quad & \text{if } \ell\neq i
\end{array}
\right. ,
\end{eqnarray}
$\bar{\mathbf{p}}_{\ell,u} = \mathbf{T}_{\mathrm{UE}}\bar{\mathbf{p}}_{\ell}$, and
\begin{eqnarray}
\nabla_{\mathbf{T}_{\mathrm{UE}}} \mathbf{p}_{\ell,u}&=&\nabla_{\mathbf{T}_{\mathrm{UE}}} \left( \left[\mathbf{e}_1,\mathbf{e}_2,\mathbf{e}_3\right]^{\mathrm{T}}\mathbf{T}_{\mathrm{UE}}\bar{\mathbf{p}}_{\ell}\right) \nonumber \\
&=& \left[\mathbf{e}_1,\mathbf{e}_2,\mathbf{e}_3\right]^{\mathrm{T}}\bar{\mathbf{p}}_{\ell}^{\odot}
\end{eqnarray}
\begin{eqnarray}
\nabla_{\mathbf{p}_i} \mathbf{p}_{\ell,u}&=&\nabla_{\mathbf{p}_i} \left( \left[\mathbf{e}_1,\mathbf{e}_2,\mathbf{e}_3\right]^{\mathrm{T}}\mathbf{T}_{\mathrm{UE}}\bar{\mathbf{p}}_{\ell}\right) \nonumber \\
&=& 
\left\{
\begin{array}{cll}
\mathbf{R}_{\mathrm{UE}}^{\mathrm{T}} & \quad & \text{if } \ell = i \\
\mathbf{0}_{3\times3}                                           & \quad & \text{if } \ell\neq i
\end{array}
\right. .
\end{eqnarray}
where the gradient with respect to $\mathbf{T}_{\mathrm{UE}}$ is calculated with the infinitesimal perturbation \cite[Sect. 7.1]{barfoot_2017}, and the operator $(\cdot)^{\odot}$ is
\begin{equation}\label{udot}
\left(\left[\bm{\xi}^{\mathrm{T}},\eta\right]^{\mathrm{T}}\right)^{\odot} = 
\begin{bmatrix}
\eta\mathbf{I}_3  & -\bm{\xi}_{\times} \\
\mathbf{0}^{\mathrm{T}} & \mathbf{0}^{\mathrm{T}}
\end{bmatrix},
\end{equation}

The gradients of $\bm{\nu}_{\ell}$ and $\mathbf{v}_{\ell}$ with respect to $\bm{\Theta}$ are given by
\begin{equation}
\nabla_{\bm{\Theta}}\bm{\nu}_{\ell}=
\begin{bmatrix}
\mathbf{e}_1^{\mathrm{T}} \\
\mathbf{e}_2^{\mathrm{T}}
\end{bmatrix}
\left(\frac{\mathbf{I}_3}{\mathbf{e}_3^{\mathrm{T}}\mathbf{p}_{\ell}} - \frac{\mathbf{p}_{\ell}\mathbf{e}_{3}^{\mathrm{T}}}{(\mathbf{e}_3^{\mathrm{T}}\mathbf{p}_{\ell})^2}\right)\nabla_{\bm{\Theta}}\mathbf{p}_{\ell},
\end{equation}
and
\begin{equation}
\nabla_{\bm{\Theta}}\mathbf{v}_{\ell}=
\begin{bmatrix}
\mathbf{e}_1^{\mathrm{T}} \\
\mathbf{e}_2^{\mathrm{T}}
\end{bmatrix}
\left(\frac{\mathbf{I}_3}{\mathbf{e}_3^{\mathrm{T}}\mathbf{p}_{\ell,u}} - \frac{\mathbf{p}_{\ell,u}\mathbf{e}_3^{\mathrm{T}}}{(\mathbf{e}_3^{\mathrm{T}}\mathbf{p}_{\ell,u})^2}\right)\nabla_{\bm{\Theta}}\mathbf{p}_{\ell,u},
\end{equation}
respectively.

\section{Generative Model for the Scatter Locations}\label{app:Generative-Model}
The scatter point locations are 
 synthesized by applying Markov Chain Monte Carlo (MCMC) sampling to the PDF of $\mathbf{p}_{\ell}$ given by \cite{Wen215,Wen21,Yu20,kulmer2019high}
\begin{equation}
    p_k(\mathbf{p}_{\ell}|\mathbf{r}_{\mathrm{UE}},\mathbf{r}_{\mathrm{BS}}) \propto 
\left\{
\begin{array}{cll}
R_k(\mathbf{p}_{\ell},\mathbf{r}_{\mathrm{UE}},\mathbf{r}_{\mathrm{BS}},\beta) & \quad & \text{if } \mathbf{p}_{\ell} \in \mathcal{S}_k \\
0        & \quad & \text{otherwise,}
\end{array}
\right. 
\end{equation}
where $\mathcal{S}_k$ for $i\in\{1,2\}$ denotes the space of the $k^{\text{th}}$ facade, and the pattern function $R_k(\mathbf{p}_{\ell},\mathbf{r}_{\mathrm{UE}},\mathbf{r}_{\mathrm{BS}},\beta)$ is \cite{Vittorio07} 
\begin{equation}\label{Rk}
    R_k(\mathbf{p}_{\ell},\mathbf{r}_{\mathrm{UE}},\mathbf{r}_{\mathrm{BS}},\beta) \propto 
\left\{
\begin{array}{cll}
\frac{\cos\theta_i\cos\theta_s}{d_i^2d_s^2} & \quad & \text{if } \beta = 0 \\
\frac{\cos\theta_i(1+\cos\psi_b)^{\beta}}{F_{\beta}d_i^2d_s^2}        & \quad & \text{otherwise.}
\end{array}
\right. 
\end{equation}
In \eqref{Rk}, $d_i = \|\mathbf{r}_{\mathrm{BS}}-\mathbf{p}_{\ell}\|$, $d_s = \|\mathbf{r}_{\mathrm{UE}}-\mathbf{p}_{\ell}\|$, the angles $\theta_i$ and $\theta_s$ are, respectively, the incidence and scattering directions with respect to $\mathbf{p}_{\ell}$, $\psi_b$ denotes the angle between the reflection and scattering directions at $\mathbf{p}_{\ell}$, $\beta \in \mathbb{N}$ describes the directivity of the scattering at $\mathbf{p}_{\ell}$, and the normalization factor $F_{\beta}$ is given by
\begin{equation}
    F_{\beta} = \frac{1}{2^{\beta}}\sum_{j=0}^{\beta}
    \left(\begin{matrix}
    \beta \\
    j
    \end{matrix}\right)\cdot I_j
\end{equation}
and
\begin{equation}
    I_j = \frac{2\pi}{j+1}\Big[\cos\theta_i \sum_{w=0}^{\frac{j-1}{2}}
    \left(\begin{matrix}
    2w \\
    w
    \end{matrix}\right)\cdot
    \frac{\sin^{2w}\theta_i}{2^{2w}}
    \Big]^{\left(\frac{1-(-1)^j}{2}\right)}.
\end{equation}
In this paper, the channel gain $\alpha_{\ell}$ of the $\ell^{\text{th}}$ path is modeled with the exponential decay model \cite{Samimi16}. More specifically, the $k^{\text{th}}$ cluster is assigned with a power of $P_k = e^{-\frac{\tau_k}{D_c}}10^{\frac{Z_k}{10}}$, 
where $Z_k \sim \mathcal{N}(0,\sigma^2_Z)$, and $\tau_k$ is the speculator delay contributed by the $k^{\text{th}}$ facade. Further, within the $k^{\text{th}}$ cluster, the $n^{\text{th}}$ scatter is associated with a channel gain of 
\begin{equation}
    \alpha_{k,n} = \frac{P'_{k,n}}{\sum_i^{N} P'_{k,i}} P_k,
\end{equation}
where
    $P'_{k,n} = e^{-\frac{\tau_{k,n}}{D_s}}10^{\frac{U_n}{10}}$,
$U_n \sim \mathcal{N}(0,\sigma^2_U)$, and $\tau_{k,n}$ is the delay associated with $n^{\text{th}}$ scatter within the $k^{\text{th}}$ facade. Following the parameters given in  \cite{Samimi16}, it is chosen such that $D_c = 25.9$~ns, $\sigma_Z = 1$~dB, $D_s = 16.9$~ns, and $\sigma_U = 6$~dB.
\ifCLASSOPTIONcaptionsoff
  \newpage
\fi
\balance 
\bibliographystyle{IEEEtran}
\bibliography{IEEEabrv,./ref}

\end{document}